\documentclass[preprint,12pt,authoryear]{elsarticle}
\usepackage{graphicx} % Required for inserting images
\usepackage{xurl}
\usepackage{amsmath}
\usepackage{amssymb}
\usepackage{amsthm} % Provides the proof environment
 \usepackage{algorithm}
\usepackage{algpseudocode}

\newtheorem{remark}{Remark}

\usepackage{hyperref}
\usepackage{xcolor}
\usepackage{caption}
\usepackage{subcaption} % Required for the subfigure environment

\newcommand{\imag}{\text{Im}}
\newcommand{\zlcomment}[1]{{\color{black} #1}}

\begin{document}
\begin{frontmatter}
\title{Constrained bilinear optimal control of reactive evolution equations}
\author{Zhexian Li$^{1}$\footnote{Corresponding author. Email: zhexianl@usc.edu}, Felipe P. J. de Barros$^1$, and Ketan Savla$^{1,2,3}$}
\address{$^1$Sonny Astani Department of Civil and Environmental Engineering, \\ University of Southern California, USA\\
$^2$Daniel J. Epstein Department of Industrial and Systems Engineering, \\ University of Southern California, USA\\
$^3$Ming Hsieh Department of Electrical and Computer Engineering,\\ University of Southern California, USA \\}

\date{\today}
\begin{abstract}
    We consider constrained bilinear optimal control of second-order linear evolution partial differential equations (PDEs) with a reaction term on the half line, where control arises as a time-dependent reaction coefficient and constraints are imposed on the state and control variables.
    These PDEs represent a wide range of physical phenomena in fluid flow, heat, and mass transfer.
    Existing computational methods for this type of control problems only consider constraints on the control variable \zlcomment{and lack global convergence guarantee}.
    In this paper, we propose a novel \zlcomment{optimize-then-discretize} framework for computing constrained bilinear optimal control with both state and control constraints. 
    % \ksmargin{"directly analyze"->"numerically solve"?}
    \zlcomment{Unlike existing methods that derive optimality conditions directly from the PDE constraint}, this framework \zlcomment{first} replaces the PDE constraint with an equivalent integral representation of the PDE solution \zlcomment{and then derives optimality conditions for the reformulated problem}.
    The integral representation, derived from the unified transform method, does not involve differential operators, and thus explicit expressions for necessary conditions of optimality can be derived using the Karush-Kuhn-Tucker conditions for infinite-dimensional optimization.
    Discretizing the optimality conditions results in a \zlcomment{system of finite-dimensional smooth nonlinear equations}, which can be efficiently solved using \zlcomment{existing algorithms with guaranteed global convergence at a quadratic rate}.
    This is in contrast with \zlcomment{discretize-then-optimize} methods that discretize the PDE first and then solve the optimality conditions of the approximated finite-dimensional problem.
    Computational results for two applications, namely nuclear reactivity control and water quality treatment in a reactor, are presented to illustrate the effectiveness of the proposed framework.
\end{abstract}
\end{frontmatter}

% \section*{Highlights (will be included as a separate file in submission)}
% \begin{itemize}
%     \item A novel optimize-then-discretize framework for computing constrained bilinear optimal control of evolution equations with a reaction term.
%     \item Deriving integral representations of PDE solutions as explicit control-to-state mappings for evaluating derivatives.
%     \item Formulating optimality conditions as infinite-dimensional equations and inequalities that can be solved by existing solvers after discretization.
%     \item Methodology is illustrated in nuclear reactivity control and water quality treatment applications.
% \end{itemize}

\section{Introduction}

% \begin{itemize}
%     \item Paragraph 1: Importance of control and optimization of physical systems governed by PDEs.
%     \item Paragraph 2 and 3: Previous works in the control community that tackled this issue: from ODEs to PDEs. Drive the reader to the challenges in applying control theory to PDEs and current state-of-the-art in this field. Bring emphasizes on mechanics (e.g. solid and fluid) and dynamical systems.
%     \item Paragraph 4: Fundamental question we wish to address? Or simply, what we wish to propose (framework? methodology?). Emphasize the novelty.
% \end{itemize}

Linear evolution PDEs are widely used to model various physical phenomena such as momentum, heat, and mass transfer in natural and engineered systems. The design of such systems to achieve the desired performance is often based on steady-state operating conditions, \zlcomment{ e.g., constant injection/production rate in subsurface reservoirs \citep{brouwer2001recovery}}. More efficient time-varying operations \zlcomment{using dynamically scheduled inputs, such as dynamic injection/production rate,} have not yet been widely used in practice, despite the fact that they have the potential to further improve performance \zlcomment{\citep{brouwer2004dynamic}}.
This is partly because these systems are infinite-dimensional due to their spatially distributed feature, which makes it hard to perform necessary computation for time-varying operations. Recently, there has been an increased interest in achieving time-varying operations of these systems \citep{haber2018sparsity,zheng2023numerical} motivated by technological advances in different fields, such as microelectromechanical systems \citep{ho1996mems}, microfluidic devices \citep{prohm2013optimal}, and smart materials, for example, used in subsurface reservoir wells \citep{brouwer2004dynamic}. 
In this work, we propose a novel framework for computing time-varying optimal control of second-order linear evolution PDEs with a reaction term.
 
Optimal control of second-order linear evolution PDEs, e.g., advection-diffusion-reaction (ADR) equations, has been studied in various applications, such as chemical process control \citep{li2008optimal} and water disinfection control \citep{elsherif2024disinfectant}.
In these references, control appears in the PDEs as either additive forcing terms or boundary conditions. 
This class of controls is called additive controls since they arise in PDEs as additive terms \citep{glowinski2022bilinear}.
Following the pioneering work of \cite{lions1971optimal}, plenty of computational methods have been developed to solve optimal control problems with additive controls, see for example \cite{troltzsch2010optimal,zuazua2015numerics} and references therein.
% Explicit optimal control expressions in the special case of linear quadratic regulator have also been considered in \cite{bamieh2002distributed,li2024complex,li2024linear}.

One of the main limitations of additive controls is that they often cannot be implemented in practice.
For example, in the case of nuclear reactivity control, an example of additive controls consists of adding or withdrawing neutrons from the nuclear reactor \citep{khapalov2010controllability}. 
For contaminated water treatment, controlling the sink term of the PDE can be interpreted as removing contaminants from the water.
These control actions are often unrealistic, as the amount of contaminants or neutrons in the system is not easily manipulated.
On the other hand, neutron absorption can be controlled by, e.g., adding or diluting chemical shim to the reactor core \citep{duderstadt1976nuclear}.
Similarly, the decay rate of contaminants in water can be controlled by using catalysts to accelerate chemical reactions \citep{heck2019catalytic}.
These changes in the principal intrinsic property of the system are usually described by controlling the coefficients in the PDEs, which are called \emph{bilinear controls} or \emph{multiplicative controls} \citep{glowinski2022bilinear,khapalov2010controllability}.

\zlcomment{Although bilinear optimal control is of practical importance, only a few studies have focused on this problem \citep{kroner2009priori} and proposed efficient optimize-then-discretize computational methods \citep[see][]{braack2018optimization,borzi2016multigrid,glowinski2022bilinear,casas2025bilinear}.
}These studies rely on abstract control-to-state operators that map the control variable to the PDE state whose derivatives can be evaluated by solving the corresponding adjoint PDE.  
Then, first-order optimality conditions are represented by a coupled system involving the original governing PDE, the corresponding adjoint PDE, and an equation in variational form forcing derivatives to be zero.
Since these PDEs depend nonlinearly on the control variable, numerically solving the coupled system of equations is challenging, and only simple control constraints have been considered, such as box constraints \citep{kroner2009priori,borzi2016multigrid,casas2025bilinear} and divergence-free constraints \citep{glowinski2022bilinear}.
To the best of authors' knowledge, \zlcomment{only \emph{local} (quadratic) convergence of the numerical methods has been established for bilinear optimal control of PDEs \citep{casas2025bilinear}}.
Moreover, no existing work has considered state constraints in bilinear optimal control problems. 
State constraints are of great practical use, e.g., to ensure that the concentration of contaminants in water does not exceed a certain safe standard or the neutron flux in a nuclear reactor does not exceed a certain operation level.
In the context of additive controls, several methods have been proposed to handle state constraints \citep{schiela2009barrier,glowinski2020admm}, but it is not obvious how to extend these methods to bilinear controls in a numerically tractable manner.
An alternative approach is to first discretize the PDE and then solve the resulting finite-dimensional optimal control problem.
In general, however, such a discretize-then-optimize approach does not necessarily provide a solution that is consistent with optimality conditions of the original infinite-dimensional optimal control problem \citep{liu2019non}.
%In general, however, this type of discretize-then-optimize approaches does not necessarily provide solutions that are consistent with optimality conditions of the original infinite-dimensional optimal control problem \citep{liu2019non}.

\zlcomment{To develop \emph{globally} converging numerical methods for bilinear optimal control with both control and state constraints, we propose an alternative optimize-then-discretize approach to deriving optimality conditions.
Instead of analyzing state-to-control operators induced by the PDE constraint as in existing studies, we replace the PDE constraint with an equivalent integral representation of the PDE solution derived from the recently developed unified transform method} \citep{fokas1997unified,fokas2008unified,deconinck2014method}. 
 This method, also known as the Fokas method, has been developed to provide an integral representation of the solution to general linear and a class of nonlinear PDEs with general boundary conditions \citep[see][]{fokas1997unified,fokas2008unified,deconinck2014method}. Among the advantages of the unified transform method is that the integral representation  converges uniformly to the given boundary conditions \citep{de2019hybrid,fokas2023modern}.
%This method, also known as the Fokas method, has been developed to provide integral representations of the solutions to general linear and a class of nonlinear PDEs with general boundary conditions \citep[see][]{fokas1997unified,fokas2008unified,deconinck2014method}.
%Among the advantages of the unified transform method, the resulting integral representations of the solutions converge uniformly to the given boundary conditions \citep{de2019hybrid,fokas2023modern}.
Therefore, the optimal control problem can be equivalently reformulated by replacing the PDE constraint with the integral representation of the PDE solution.
This is not necessarily true for traditional methods for solving PDEs, such as separation of variables. 
For example, PDE solutions represented by sine transforms and series converge to zero at the boundary $x=0$, which is consistent only with homogeneous boundary conditions \citep{olver2014introduction}. 
Moreover, integral representations derived from the unified transform method can be seen as functionals of the control variable, which allows the derivation of explicit expressions for directional derivatives of the objective and constraints with respect to the control variable.
Then, necessary conditions of optimality can be derived from the celebrated Karush-Kuhn-Tucker (KKT) conditions in function spaces \citep{hinze2008optimization,attouch2014variational}, in principle, for any objectives and constraints that are differentiable in appropriate sense. 
The unified transform method has been applied mainly to solve PDEs with constant coefficients \citep{deconinck2014method,fokas2023modern} and related additive control problems \citep{kalimeris2023numerical,li2024complex,li2024linear}. In this paper, we also extend this method to PDEs with a time-varying reaction coefficient and the corresponding constrained bilinear optimal control problem.

\zlcomment{Our main contribution is an \emph{optimize-then-discretize} framework for constrained bilinear optimal control using the unified transform method. This method provides an exact integral representation of the PDE solution, which we leverage to replace the PDE constraint within the control problem. The resulting optimality conditions form a system of smooth nonlinear equations that can be directly solved by existing algorithms with guaranteed global convergence at a quadratic rate.} 
The framework introduced in this paper consists of the following steps.
\begin{enumerate}
    \item[\zlcomment{1.}] For a given set of initial and boundary conditions, we derive an integral representation of the solution using the unified transform method.
Replacing the PDE constraint with the integral representation results in an equivalent reformulation of the optimal control problem.
\item[\zlcomment{2.}] For the reformulated problem after replacing the PDE constraint, we derive explicit expressions for the directional derivatives of the objective and constraints with respect to the control variable.
These directional derivatives are used in necessary conditions of optimality derived from the KKT conditions.
The optimality conditions are in the form of a system of infinite-dimensional equations and inequalities. 
\item[\zlcomment{3.}]  We discretize and reformulate the optimality conditions as a system of finite-dimensional smooth nonlinear equations.
Unlike discretize-then-optimize approaches that first discretize the PDE, the discretization of the optimality conditions in our framework does not involve differential operators and thus preserves the PDE relation in continuous space and time.
\end{enumerate}

This paper is organized as follows. 
Section \ref{sec:problem} formulates the constrained bilinear optimal control problem for second-order linear evolution equations and presents two applications of the control problem.
The computational framework for solving the control problem is described in Section \ref{sec:computation-framework}.
The computational results for the two applications are presented in Section \ref{sec:results}.
Finally, we provide concluding remarks in Section \ref{sec:conclusion}.

% \section{Literature Review}
% \begin{enumerate}
%     \item \citet{glowinski2022bilinear} is the state-of-the-art for bilinear optimal control of advection-diffusion systems. The authors derived first-order optimality conditions and claimed to provide the first efficient numerical algorithms to obtain optimal solutions.
%     \item \citet{wang2018free} provided a numerical approach and convergence analysis to the ensemble bilinear optimal control problem with only running cost on control and terminal cost on state.
% \end{enumerate}

\section{Problem Statement}
\label{sec:problem}
In this section, we formulate the constrained bilinear optimal control problem for second-order linear evolution equations with a reaction term.
Then, we specify two applications of the control problem for nuclear reactors and solute transport in fluids, which will be investigated in this work.

\subsection{Constrained bilinear optimal control problem}
\label{sec:problem-formulation}
This paper considers second-order linear evolution equations with a reaction term defined on the half line $x\geq0$ and finite time interval $0\leq t\leq T$.
Let $\psi(x,t)$ denote the state variable, $v$ denote the constant velocity, $\alpha$ denote the constant diffusion coefficient, $u(t)$ denote the time-varying reaction coefficient, and $f(x,t)$ denote the source term.
The PDE considered in this paper is of the form
\begin{equation}\label{eq:general-second-order}
    \frac{\partial \psi(x,t)}{\partial t} + v \frac{\partial \psi(x,t)}{\partial x} = \alpha \frac{\partial^2 \psi(x,t)}{\partial x^2} - u(t)\psi(x,t) + f(x,t),
\end{equation}
with appropriate initial and boundary conditions specified according to the application. 
% \begin{align}
%     \psi(x,0) &= \psi_o(x),\quad x>0,\label{eq:ic-general}\\
%     \alpha\psi(0,t) + \beta\psi_x(0,t) &= g(t),\quad 0<t<T\label{eq:bc-general} \\
%     \lim_{x\to\infty}\psi(x,t) &= 0, \quad 0<t<T\label{eq:bc-infinity}
% \end{align}
% where $\psi_o(x)$ and $g(t)$ are given functions, $\alpha,\beta$ are constant coefficients corresponding to Dirichlet and Neumann boundary values at $x=0$, respectively.
The PDE \eqref{eq:general-second-order} can be used to describe a wide range of physical phenomena, such as the convection and diffusion of heat and transport of solutes in fluids, see \cite{mikhailov1984unified,cotta1993integral} and references therein.
\zlcomment{The existence, uniqueness, and regularity of the solution to \eqref{eq:general-second-order} is well-established in standard PDE analysis, see for example \citet[Chapter 7.1]{evans2022partial}. 
In this paper, we restrict our attention to the space of square-integrable functions, namely coefficient $u(t)$, source $f(x,t)$, initial, and boundary conditions are all square-integrable functions. 
This is consistent with the setting in \cite{glowinski2022bilinear} and sufficient to apply existing results on the existence of optimal control \citep[Theorem 1.45]{hinze2008optimization}.}

We are interested in controlling the reaction coefficient $u(t)$ in \eqref{eq:general-second-order} to optimize the desired performance of the system. 
% In practice, we often consider a parameterized control $u(t;\theta)$ where $\theta\in\mathbb{R}^m$ is a vector of parameters.
% For example, the control $u(t)$ may not be able to change at every time $t$. If we divide the time interval $[0,T]$ into $m$ subintervals $[t_i,t_{i+1}],i=1,\ldots,m$, we can define the control as a piecewise constant function, i.e., $u(t;\theta)$ = $\theta_i$ if $t\in[t_i,t_{i+1}]$.
% Other parameterization of the control is also possible, such as polynomial or truncated Fourier series.
For performance metrics, we consider the following objective functional,
\begin{equation}\label{eq:objective-functional}
    \int_{0}^{T}\int_{0}^{\infty}w_1\psi^2(x,t) dx\ dt + \int_0^T w_2 u^2(t) dt,
\end{equation} 
where $w_1$ and $w_2$ are nonnegative coefficients that weigh the state and the control, respectively.
We also consider the following constraints on the state and control:
\begin{align}
    \psi_{\min}(x,t)&\leq \psi(x,t) \leq \psi_{\max}(x,t),\quad L_1\leq x\leq L_2, 0\leq t\leq T, \label{eq:state-constraint-general} \\
    u_{\min}(t)&\leq u(t) \leq u_{\max}(t), \quad 0\leq t\leq T, \label{eq:control-constraint-general}
\end{align}
where $\psi_{\min}(x,t)$ and $\psi_{\max}(x,t)$ are the given nonnegative upper and lower bounds of the state, $u_{\min}(t)$ and $u_{\max}(t)$ are the given nonnegative upper and lower bounds of the control, $L_1$ and $L_2$ specify the space interval where the state constraint \eqref{eq:state-constraint-general} is added.
The optimal control problem is to find an optimal $u^*$ that minimizes the objective in \eqref{eq:objective-functional} subject to the PDE \eqref{eq:general-second-order}, the initial and boundary conditions, and the constraints \eqref{eq:state-constraint-general} and \eqref{eq:control-constraint-general}.
\zlcomment{The existence of optimal control $u^*$ follows from \citet[Theorem 1.45]{hinze2008optimization} provided that the feasible set determined by constraints \eqref{eq:state-constraint-general} and \eqref{eq:control-constraint-general} is nonempty.
In practice, the upper and lower bounds in \eqref{eq:state-constraint-general} and \eqref{eq:control-constraint-general} are determined by physical requirements of the system and limitations of the control, which shall be properly chosen such that the feasible set is nonempty.}

The control $u(t)$ appears as the reaction coefficient in \eqref{eq:general-second-order}, which belongs to the class of bilinear controls.
This type of control problem has only been solved without state constraints, see for example \cite{borzi2016multigrid,glowinski2022bilinear}.
Moreover, their approaches involve discretizing the PDE in space and time. 
In this paper, we reformulate the problem by replacing the PDE constraint with an integral representation of the solution derived from the unified transform method \citep{fokas1997unified,deconinck2014method}.
% The unified transform method can be used to derive integral representations of the solutions to linear PDEs, see for example \cite{fokas1997unified,deconinck2014method,de2019hybrid}.
The integral representation does not involve differential operators.
The advantage of our approach is that we can handle state and control constraints such as \eqref{eq:state-constraint-general} and \eqref{eq:control-constraint-general} and avoid discretizing the PDE. 
In Sections \ref{sec:nuclear-reactor} and \ref{sec:ADE}, we present two applications of the constrained bilinear optimal control problem and the corresponding integral representations of the solutions.

\zlcomment{
\begin{remark}
    The coefficients $v$ and $\alpha$ in \eqref{eq:general-second-order} are assumed to be constant in this paper.
    However, our method can be extended to the case where $v(x)$ and $\alpha(x)$ are piecewise constant spatially varying coefficients, for which the unified transform method can still be applied following \cite{deconinck2014method}.

    As for the control coefficient, our method can be applied to either time-varying control $u(t)$ used in \eqref{eq:general-second-order}, or spatially varying control $u(x)$ following the unified transform method illustrated in \cite{fokas2004boundary} -- the latter will not be discussed in this paper.
    The time-varying control is more relevant to the applications considered in this paper.
    For example, in the water treatment problem presented in Section~\ref{sec:ADE}, it is easier to change the reaction coefficient uniformly in space over time than the spatial distribution of the reaction coefficient by adding different amounts of catalysts to water.
\end{remark}
}
\begin{remark}
Although we only consider box constraints in \eqref{eq:state-constraint-general} and \eqref{eq:control-constraint-general} that are most relevant to our application problems, the proposed framework can be applied to more general constraints that are differentiable in appropriate sense, such as constraints that are differentiable functions of the state $\psi$ and control $u$.
\end{remark}

\subsection{Nuclear reactivity control}
\label{eq:nuclear-reactor}
In this example, we consider the case of nuclear reactivity, which is relevant for applications in nuclear physics \cite[Chapter 5.II]{duderstadt1976nuclear}. For the purpose of illustration, we consider a one-dimensional system in which neutrons are uniformly emitted from a planar source. Neutron transport is assumed to be diffusive within an infinite homogeneous medium in the presence of absorption. The neutron source is assumed to emit neutrons isotropically at a rate of $S_0(\tilde{t})$ [L$^{-2}$T$^{-1}$] at time $\tilde{t}$ [T]. In this model, we assume that all neutrons travel at the same speed $v_n$ [L/T]. Following the work of \citet{duderstadt1976nuclear}, let
$\tilde{\phi}(\tilde{x},\tilde{t})$ [L$^{-2}$T$^{-1}$] denote the neutron flux at a given longitudinal distance $\tilde{x}$ [L] from the source.  The neutron flux can be expressed as $\tilde{\phi}(\tilde{x},\tilde{t})=v_nN(\tilde{x},\tilde{t})$ where $N$ [L$^{-3}$] represents the neutron density.
Let $D_n$ denote the diffusivity coefficient [L], and $\Sigma_a(\tilde{t})$ denote the macroscopic absorption coefficient [L$^{-1}$] at time $\tilde{t}$\zlcomment{, i.e.,} the probability of neutron absorption in a macroscopic sample of the medium.

The evolution of neutron flux can be described by the one-speed neutron diffusion equation \citep{duderstadt1976nuclear,espinosa2008constitutive}:
\begin{equation}\label{eq:diffusion-equation-neutron}
    \frac{1}{v_n}\frac{\partial \tilde{\phi}(\tilde{x},\tilde{t})}{\partial \tilde{t}} = D_n\frac{\partial^2 \tilde{\phi}(\tilde{x},\tilde{t})}{\partial \tilde{x}^2} - \Sigma_a(\tilde{t}) \tilde{\phi}(\tilde{x},\tilde{t}),
\end{equation}
with the initial and boundary conditions given, respectively, by
\begin{align*}
\tilde{\phi}(\tilde{x},0)&=\tilde{\phi}_o(\tilde{x}),\quad \tilde{x}\geq0,\\
-D_n\frac{\partial\tilde{\phi}(0,\tilde{t})}{\partial\tilde{x}}&=\frac{S_0(\tilde{t})}{2}, \quad 0\leq\tilde{t}\leq\tilde{T},\\
\lim_{\tilde{x}\to\infty}\tilde{\phi}(\tilde{x},t)&=0, \quad 0\leq\tilde{t}\leq\tilde{T}.    
\end{align*}
\zlcomment{
\begin{remark}
 The zero boundary condition at infinity is standard in many areas, e.g., heat conduction \citep{hahn2012heat} and solute mass transport \cite{lee2019applied}, and nonzero constant boundary conditions can be transformed to zero boundary conditions using a change of variables. 
\end{remark}
} 
Following \citet[Chapter 14.IV]{duderstadt1976nuclear}, controlling the macroscopic cross-section $\Sigma_a$ in \eqref{eq:diffusion-equation-neutron} can be accomplished by adding or diluting chemical shim, e.g., boric acid, in the reactor core. 
The chemical shim has high neutron absorption cross-section and is typically dissolved homogeneously in the coolant in the entire reactor core.

To rewrite \eqref{eq:diffusion-equation-neutron} in a dimensionless form, we introduce the following dimensionless quantities:
\begin{equation}\label{eq:dimensionless-variables-neutron}
    \phi = \frac{\tilde{\phi}}{\bar{\phi}}, \quad x = \frac{\tilde{x}}{D_n}, \quad t = \frac{v_n\tilde{t}}{D_n},
\end{equation}
where $\bar{\phi}$ is a characteristic neutron flux value.
The dimensionless equation can now be expressed as
\begin{equation}\label{eq:dimensionless-diffusion-equation-neutron}
    \frac{\partial \phi(x,t)}{\partial t} = \frac{1}{D_n}\frac{\partial^2 \phi(x,t)}{\partial x^2} - D_n\Sigma_a(t)\phi(x,t),
\end{equation}
with the initial condition $\phi(x,0)=\phi_o(x) = \tilde{\phi}_o(x) / \bar{\phi}$ and the Neumann boundary condition $$\frac{\partial \phi(0,t)}{\partial x}=\xi_{\text{Ne}}(t) = -\frac{S_0(t)}{2\bar{\phi}}.$$
Note that \eqref{eq:dimensionless-diffusion-equation-neutron} can be written in the form \eqref{eq:general-second-order} with $v=0$, $\alpha=1/D_n$, $u(t)=D_n\Sigma_a(t)$, and $f(x,t)=0$.

Using the unified transform method \citep{fokas1997unified}, an integral representation of the solution to \eqref{eq:dimensionless-diffusion-equation-neutron} can be written as
\begin{multline}\label{eq:integral-representation-final-neutron}
    \phi(x,t,\Sigma_a) = \int_{-\infty}^{\infty} e^{ikx-\omega_\phi(k,t,\Sigma_a)}\hat{\phi}_o(k) \frac{dk}{2\pi} \\
     + \int_{\partial\mathcal{D}^+} e^{ikx-\omega_\phi(k,t,\Sigma_a)}\left[\hat{\phi}_o\left(-k\right) - \frac{2}{D_n}\hat{\xi}_{\text{Ne}}(k,t,\Sigma_a)\right]\frac{dk}{2\pi},
\end{multline}
where $i$ is the imaginary unit, $\partial\mathcal{D}^+ = \{k\in\mathbb{C}^+:k=|k|e^{i\theta},\theta = \pi / 8 \text{ or }7\pi / 8\}$ is the union of two line segments in the upper-half complex plane with arguments of $\pi/8$ and $7\pi/8$, respectively, and 
\zlcomment{
\begin{equation}\label{eq:global-relation-terms-neutron}
\begin{split}
    &\omega_\phi(k,t,\Sigma_a) = k^2t/D_n +D_n\Sigma_a^{\text{int}}(t), \quad \Sigma_a^{\text{int}}(t) = \int_{0}^{t} \Sigma_a(\tau)d\tau,\\
    &\hat{\phi}_o(k) = \int_{0}^{\infty} \phi_o(x)e^{-ikx}dx, \quad
    \hat{\xi}_{\text{Ne}}(k,t,\Sigma_a) = \int_{0}^{t} e^{\omega_\phi(k,\tau,\Sigma_a)}\xi_{\text{Ne}}(\tau)d\tau.
\end{split}
\end{equation}
}Details of the derivation of \eqref{eq:integral-representation-final-neutron}--\eqref{eq:global-relation-terms-neutron} are presented in \ref{sec:pde-representation}.
We emphasize the dependence of $\phi$ on the control variable $\Sigma_a$ in the left-hand side of \eqref{eq:integral-representation-final-neutron}.
The integral representation \eqref{eq:integral-representation-final-neutron} can be used to replace the PDE constraint in the constrained bilinear optimal control problem formulated in Section~\ref{sec:problem-formulation}.

\label{sec:nuclear-reactor}

\subsection{Solute transport in fluids}
\label{sec:ADE}

Next, we consider advective and dispersive transport of a reactive solute, e.g., a pollutant, within a channel. For the sake of illustration, flow and transport are assumed to be one-dimensional. Such problems are of relevance to solute transport phenomena in porous media \citep{lee2019applied}, pipelines \citep{shang2021lagrangian} and rivers \citep{de2007integral,genuchten2013exact}.
%In this subsection, we consider solute transport in fluids following %\cite{lee2019applied}. 
Let $\tilde{C}(\tilde{x},\tilde{t})$ [M/L$^{3}$] represent the concentration of a dissolved solute at position $\tilde{x}$ [L] and time $\tilde{t}$ [T]. The longitudinal dispersion coefficient is given by $D_c$ [L$^2$/T] and $v_c$ [L/T] denotes the velocity of the fluid (assumed to be uniform). The first-order decay rate is given by $\lambda_c(\tilde{t})$ [1/T]. For our problem, we consider
a time-varying concentration $V_0(\tilde{t})$ at the inlet location of the computational domain.
The spatial temporal evolution of the solute concentration is governed by the following PDE
\begin{equation}\label{eq:solute-transport-equation}
    \frac{\partial \tilde{C}(\tilde{x},\tilde{t})}{\partial \tilde{t}} + v_c \frac{\partial \tilde{C}(\tilde{x},\tilde{t})}{\partial \tilde{x}} = D_c \frac{\partial^2 \tilde{C}(\tilde{x},\tilde{t})}{\partial \tilde{x}^2} - \lambda_c(\tilde{t})\tilde{C}(\tilde{x},\tilde{t}),
\end{equation}
with the initial and boundary conditions given, respectively, by
\begin{align*}
\tilde{C}(\tilde{x},0)&=\tilde{C}_o(\tilde{x}),\quad \tilde{x}\geq0,\\
\tilde{C}(0,\tilde{t})&=V_0(\tilde{t}), \quad 0\leq\tilde{t}\leq\tilde{T},\\
\lim_{\tilde{x}\to\infty}\tilde{C}(\tilde{x},t)&=0, \quad 0\leq\tilde{t}\leq\tilde{T}.
\end{align*}
The decay rate $\lambda_c$ is usually determined by the chemical reaction rate between the solute and other reactants injected in the ambient fluids, see for example chlorine decay in water distribution systems \citep{powell2000factors,hallam2002decay}.
Therefore, controlling the decay rate \zlcomment{$\lambda_c(\cdot)$} of the solute plume can be realized using catalysts to change the chemical reaction rate.
For example, in the case of water treatment, the reaction rate of toxic contaminants such as chlorinated organics and nitrates can be accelerated using a catalyst converter for water \citep{heck2019catalytic}.

The following dimensionless quantities are adopted to rewrite \eqref{eq:solute-transport-equation} in a dimensionless form,
\begin{equation}
    C = \frac{\tilde{C}}{\bar{C}}, \quad x = \dfrac{v_c\tilde{x}}{D_c}, \quad t= \dfrac{v_c^2\tilde{t}}{D_c},
\end{equation}
where $\bar{C}$ is a characteristic solute concentration value at the inlet boundary.
The resulting dimensionless equation is
\begin{equation}\label{eq:dimensionless-solute-transport-equation}
    \frac{\partial C(x,t)}{\partial t} + \frac{\partial C(x,t)}{\partial x} = \frac{\partial^2 C(x,t)}{\partial x^2} - \frac{D_c\lambda_c(t)}{v_c^2}C(x,t),
\end{equation}
with the initial condition $C(x,0)=C_o(x) = \tilde{C}_o(x) / \bar{C}$ and the Dirichlet boundary condition $C(0,t)=C_{\text{Di}}(t) = V_0(t) / \bar{C}$.
Note that \eqref{eq:dimensionless-solute-transport-equation} is in the form \eqref{eq:general-second-order} with $v=1$, $\alpha=1$, $u(t)=D_c\lambda_c(t)/v_c^2$, and $f(x,t)=0$.
Using the unified transform method, an integral representation of the solution to \eqref{eq:dimensionless-solute-transport-equation} is given by

\begin{multline}\label{eq:integral-representation-final-solute}
    C(x,t,\lambda_c) = \int_{-\infty}^{\infty} e^{ikx - \omega_c(k,t,\lambda_c)} \hat{C}_o(k)\frac{dk}{2\pi} \\ 
    + \int_{\partial\mathcal{D}^+} e^{ikx-\omega_c(k,t,\lambda_c)}\left[- \hat{C}_o\left(-k-i\right) -\left(2ik-1\right)\hat{C}_{\text{Di}}(k,t,\lambda_c)\right]\frac{dk}{2\pi},
\end{multline}
where $\partial\mathcal{D}^+$ is defined as in \eqref{eq:integral-representation-final-neutron}, and
\zlcomment{
\begin{equation}\label{eq:global-relation-terms-solute}
\begin{split}
    &\omega_c(k,t,\lambda_c) = (k^2+ik)t +D_c\lambda_c^{\text{int}}(t)/v_c^2, \quad \lambda_c^{\text{int}}(t) = \int_{0}^{t} \lambda_c(\tau)d\tau,\\
    &\hat{C}_o(k) = \int_{0}^{\infty} C_o(x)e^{-ikx}dx, \quad
    \hat{C}_{\text{Di}}(k,t,\lambda_c) = \int_{0}^{t} e^{\omega_c(k,\tau,\lambda_c)}C_{\text{Di}}(\tau)d\tau.
\end{split}
\end{equation}
}Details of the derivation of \eqref{eq:integral-representation-final-solute}--\eqref{eq:global-relation-terms-solute} are presented in \ref{sec:pde-representation}.
We emphasize the dependence of $C$ on the control variable $\lambda_c$ on the left-hand side of \eqref{eq:integral-representation-final-solute}.
Similar to \eqref{eq:integral-representation-final-neutron}, the integral representation \eqref{eq:integral-representation-final-solute} can be used to replace the PDE constraint in the constrained bilinear optimal control problem formulated in Section~\ref{sec:problem-formulation}.

\section{Computational framework}
\label{sec:computation-framework}
In this section, we present a computational framework for solving the constrained bilinear optimal control problem \eqref{eq:objective-functional}--\eqref{eq:control-constraint-general}.
The framework consists of the following steps:
\begin{enumerate}
    \item[\textit{Step 1.}] Reformulation of the constrained bilinear optimal control problem by replacing the PDE constraint with an integral representation of the solution derived from the unified transform method, e.g., replacing \eqref{eq:dimensionless-diffusion-equation-neutron} with \eqref{eq:integral-representation-final-neutron} or \eqref{eq:dimensionless-solute-transport-equation} with \eqref{eq:integral-representation-final-solute}.
    \item[\textit{Step 2.}] Derivation of  optimality conditions as a system of infinite-dimensional equations and inequalities for the reformulated problem using the KKT conditions.
    \item[\textit{Step 3.}] Discretizing the optimality conditions to obtain a finite-dimensional system of nonlinear equations that can be solved numerically through existing solvers such as \texttt{fsolve} in MATLAB.
\end{enumerate}
In the followings, we present the details of each step.

\subsection*{Step 1: Reformulation of the constrained bilinear optimal control problem}
Following \eqref{eq:integral-representation-final-neutron} and \eqref{eq:integral-representation-final-solute},
let $\psi(x,t,u)$ denote a given integral representation of the solution to \eqref{eq:general-second-order}.
Then, the constrained bilinear optimal control problem can be equivalently reformulated as
\begin{equation}
    \label{eq:optimal-control-reformulate}
    \begin{split}
        \min_{u} &\quad J(u) = \int_{0}^{T}\int_{0}^{\infty}w_1\psi^2(x,t,u) dx\ dt + \int_0^T w_2 u^2(t) dt \\
        \text{s.t. } & \psi_{\min}(x,t)\leq \psi(x,t,u) \leq \psi_{\max}(x,t),\quad L_1\leq x\leq L_2, 0\leq t\leq T, \\
    &u_{\min}(t)\leq u(t) \leq u_{\max}(t), \quad 0\leq t\leq T.
    \end{split}
\end{equation} 
The main difference between \eqref{eq:control-constraint-general} and \eqref{eq:optimal-control-reformulate} is that the differential operators $\frac{\partial \psi}{\partial x}, \frac{\partial^2 \psi}{\partial x^2}$ and $\frac{\partial \psi}{\partial t}$ do not appear in the reformulated problem \eqref{eq:optimal-control-reformulate} since $\psi(x,t,u)$ is given in an integral form.
% In Section~\ref{sec:kkt-finite-to-infinite}, we derive necessary conditions of optimality for \eqref{eq:optimal-control-reformulate} using the celebrated Karush-Kuhn-Tucker (KKT) conditions \citep{karush1939minima,kuhn2013nonlinear}.
% In Section~\ref{sec:computations-kkt}, we present a discretized version of the KKT conditions that can be solved numerically via solvers such as \texttt{fsolve} in MATLAB.

\subsection*{Step 2: Derivation of optimality conditions}
\label{sec:kkt-finite-to-infinite}
Before we derive necessary conditions of optimality for the reformulated problem \eqref{eq:optimal-control-reformulate}, we first review the classical optimality conditions, i.e., the KKT conditions \citep{karush1939minima,kuhn2013nonlinear} for finite-dimensional nonlinear optimization problems.
Consider an optimization problem of the form
\begin{equation*}
    \label{eq:optimal-control-finite-dim}
    \begin{split}
        &\min_{\pmb{u}\in\mathbb{R}^N}\quad J(\pmb{u}) \\
        &\text{s.t.}\quad  g_j(\pmb{u})\leq 0,\, j=1,\ldots,p,\\
    \end{split}
\end{equation*}
where $J:\mathbb{R}^N\to\mathbb{R}$ and $g_j:\mathbb{R}^N\to\mathbb{R}, j=1,\ldots,p$ are continuously differentiable functions.
Suppose a constraint qualification condition holds, e.g., there exists $\bar{\pmb{u}}$ such that $g_j(\bar{\pmb{u}})<0,j=1,\ldots,p$. Then, for every optimal solution $\pmb{u}^*$, there exists a Lagrangian multiplier $\pmb{\lambda}^* = [\lambda_1^*,\ldots,\lambda_p^*]$ with each $\lambda_j^*$ associated with a constraint $g_j,j=1\ldots,p$, such that the following KKT conditions hold,
\begin{equation}\label{eq:kkt-finite-dim}
    \begin{split}
    &\nabla_{\pmb{u}}\mathcal{L}(\pmb{u}^*,\pmb{\lambda}^*) = 0,\\
     & g_j(\pmb{u}^*)\leq 0, \quad \lambda_j^*\geq 0, \quad \lambda_j^* g_j(\pmb{u}^*) = 0,\quad j=1,\ldots,p,
    \end{split}
\end{equation}
where
\begin{equation}
    \mathcal{L}(\pmb{u},\pmb{\lambda}) = J(\pmb{u}) + \sum_{j=1}^{p} \lambda_j g_j(\pmb{u})
\end{equation}
is the Lagrangian function, and $\nabla_{\pmb{u}} \mathcal{L}$ is the gradient of $\mathcal{L}$ with respect to $\pmb{u}$.

There is an inherent relation between the gradient and directional derivatives, i.e., for every direction $\pmb{h}\in\mathbb{R}^N$, the directional derivative of $\mathcal{L}$ in the direction $\pmb{h}$ at $\pmb{u}$ is given by
\begin{equation}
    d_{\pmb{u}} \mathcal{L}(\pmb{u},\pmb{\lambda};\pmb{h}) = \lim_{\epsilon\to0^+}\frac{\mathcal{L}(\pmb{u}+\epsilon \pmb{h},\pmb{\lambda}) - \mathcal{L}(\pmb{u},\pmb{\lambda})}{\epsilon} = \nabla_{\pmb{u}} \mathcal{L}(\pmb{u},\pmb{\lambda}) \cdot \pmb{h},
\end{equation}
where the gradient $\nabla_{\pmb{u}} \mathcal{L}(\pmb{u},\pmb{\lambda})$ is independent of the direction $\pmb{h}$.
In other words, the gradient $\nabla_{\pmb{u}} \mathcal{L}$ consists of the directional derivatives of $J$ in the directions $\pmb{e}^j,j=1,\ldots,N$, where $\pmb{e}^j$ is the $j$-th unit vector in $\mathbb{R}^N$, i.e., the partial derivative of $\mathcal{L}$ with respect to each entry in $\pmb{u}$.

Consider the infinite-dimensional case where $u(t)$ is a scalar function defined on $0\leq t\leq T$. 
An intuitive way to extend the concept of gradient or (partial) derivative is to consider directional derivative in the direction $\delta(t-\tau),0\leq\tau\leq T$, where $\delta(\cdot)$ is the Dirac delta function. 
The Dirac delta function plays a similar role as the unit vectors in $\mathbb{R}^N$ in the sense that $\delta(t-\tau)=0$ if $t\neq \tau$.
This derivative can be written in the form 
\begin{equation}\label{eq:functional-derivative-lagrangian}
    \frac{\delta\mathcal{L}}{\delta u}(u,\pmb{\lambda},\tau) = \lim_{\epsilon\to0^+}\frac{\mathcal{L}(u+\epsilon\delta(t-\tau),\pmb{\lambda}) - \mathcal{L}(u,\pmb{\lambda})}{\epsilon},\quad 0\leq \tau\leq T.
\end{equation} 
\begin{remark}
    The derivative given by \eqref{eq:functional-derivative-lagrangian} is sometimes referred to as the functional derivative \citep{greiner2013field}.
    The limit in \eqref{eq:functional-derivative-lagrangian} is usually not defined in the sense that  $\delta(\cdot)$ may not be a valid direction, e.g., in the space of \zlcomment{square-integrable} functions.
    Nevertheless, it suffices for our purpose to provide an intuitive extension of the gradient to the infinite-dimensional case.
    A formal definition of derivatives is provided in \ref{sec:appendix-directional-derivative}.
\end{remark}

Following \eqref{eq:optimal-control-reformulate}, consider the case where the constraints $g_j(x,t,u)$ are defined on $(x,t)\in\Omega_j,j=1,\ldots,p$, and each $\Omega_j$ is a compact subset of independent variables $(x,t)$ within the computational domain $x\geq0$ and $0\leq t\leq T$. 
Let each constraint $g_j$ be associated with a Lagrange multiplier \emph{function} $\lambda_j$ defined on $\Omega_j$, and let $\pmb{\lambda} = [\lambda_1,\ldots,\lambda_j]$ be the compact notation. 
A natural extension of \eqref{eq:kkt-finite-dim} for the infinite-dimensional case are the following conditions:
\begin{equation}\label{eq:kkt-infinite-dim}
    \begin{split}
    &\frac{\delta\mathcal{L}}{\delta u}(u^*,\pmb{\lambda}^*,\tau) = 0, \quad \zlcomment{\text{for almost every }\tau\in[0,T]},\\
     & g_j(x,t,u^*)\leq 0, \quad \lambda_j^*(x,t)\geq 0,  \quad (x,t)\in\Omega_j, \quad j=1,\ldots, p,\\
     & G_j(u^*,\lambda_j^*) = 0,\quad j=1,\ldots,p,
    \end{split}
\end{equation}
where 
\begin{equation}
    \begin{split}
        G_j(u,\lambda_j) &= \iint_{\Omega_j} \lambda_j(x,t) g_j(x,t,u)dx \ dt = 0, \\
        \mathcal{L}(u,\pmb{\lambda}) &= J(u) + \sum_{j=1}^{p}G_j(u,\lambda_j).
    \end{split}
\end{equation}

For optimal control of \eqref{eq:dimensionless-solute-transport-equation} and \eqref{eq:dimensionless-diffusion-equation-neutron},
necessary conditions of optimality can indeed be written in the form \eqref{eq:kkt-infinite-dim}.
Formally, suppose that there exists $\bar{u}$ that lies in the interior of the feasible set of \eqref{eq:optimal-control-reformulate}, i.e., $\psi_{\min}(x,t)<\psi(x,t,\bar{u})<\psi_{\max}(x,t),L_1\leq x\leq L_2,0\leq t\leq T$ and $u_{\min}(t)<\bar{u}(t)<u_{\max}(t), 0\leq t\leq T$.
Then, for every optimal $u^*$ of \eqref{eq:optimal-control-reformulate}, there exists Lagrangian multiplier functions $\pmb{\lambda}^* = [\lambda_1^*,\ldots,\lambda_4^*]$ such that \eqref{eq:kkt-infinite-dim} holds for all $\tau\in[0,T]$ with
\begin{equation}
    \label{eq:kkt-infinite-lagrangian}
    \begin{split}
        &\frac{\delta\mathcal{L}}{\delta u}(u,\pmb{\lambda},\tau) = \frac{\delta J}{\delta u}(u,\tau) + \sum_{j=1}^{4} \frac{\delta G_j}{\delta u}(u,\lambda_j,\tau), \\
        &\frac{\delta J}{\delta u}(u,\tau)  =  2w_1\int_{\tau}^{T} \left[\int_{0}^{\infty}\psi(x,t,u) \frac{\delta\psi}{\delta u} (x,t,u,\tau) dx\right] dt + 2w_2u(\tau), \\
        &\frac{\delta G_j}{\delta u}(u,\lambda_j,\tau) =  (-1)^j\int_{\tau}^{T} \int_{L_1}^{L_2} \lambda_j(x,t)\frac{\delta \psi}{\delta u}(x,t,u,\tau)dx\ dt, \quad j=1,2, \\
        &\frac{\delta G_j}{\delta u}(u,\lambda_j,\tau) =  (-1)^j\lambda_j(\tau), \quad j=3,4, \\
    \end{split}
\end{equation}
and constraints 
\begin{equation}
    \label{eq:kkt-infinite-constraints}
    \begin{split}
    &g_1(x,t,u) = \psi_{\min}(x,t) - \psi(x,t,u),\quad g_2(x,t,u) = \psi(x,t,u) - \psi_{\max}(x,t),\\
        &g_3(x,t,u) = u_{\min}(t) - u(t),\quad g_4(x,t,u) = u(t) - u_{\max}(t), \\
        &\Omega_1 = \Omega_2 = \{x,t: L_1\leq x\leq L_2, 0\leq t\leq T\}, \quad \Omega_3 = \Omega_4 = \{t: 0\leq t\leq T\}.
    \end{split}
\end{equation}
For the problems specified in Sections~\ref{sec:nuclear-reactor} and \ref{sec:ADE}, see \eqref{eq:dimensionless-diffusion-equation-neutron} and \eqref{eq:dimensionless-solute-transport-equation}, $\frac{\delta\psi}{\delta u}$ is given, respectively, by 
\begin{equation}
    \label{eq:derivative-psi}
    \begin{split}
    \frac{\delta\phi}{\delta \Sigma_a}&(x,t,\Sigma_a,\tau) = - \int_{-\infty}^{\infty} e^{ikx - \omega_\phi(k,t,\Sigma_a)}D_n\hat{\phi}_o(k) \frac{dk}{2\pi} \\ & - \int_{\partial\mathcal{D}^+} e^{ikx - \omega_\phi(k,t,\Sigma_a)} D_n\left[\hat{\phi}_o(-k)  - \frac{2}{D_n} \hat{\xi}_{\text{Ne}}(k,\tau,\Sigma_a) \right] \frac{dk}{2\pi}, \\
    \frac{\delta C}{\delta \lambda_c}&(x,t,\lambda_c,\tau) = - \int_{-\infty}^{\infty} e^{ikx - \omega_c(k,t,\lambda_c)}\frac{D_c}{v_c^2}\hat{C}_o(k) \frac{dk}{2\pi} \\ &  - \int_{\partial\mathcal{D}^+} e^{ikx - \omega_c(k,t,\lambda_c)}\frac{D_c}{v_c^2}\left[- \hat{C}_o(-k-i) - (2ik-1)\hat{C}_{\text{Di}}(k,\tau,\lambda_c)\right]\frac{dk}{2\pi},
    \end{split}
\end{equation}
where $\partial\mathcal{D}^+$ is defined as in \eqref{eq:integral-representation-final-neutron} and \eqref{eq:integral-representation-final-solute}.
A formal derivation of \eqref{eq:kkt-infinite-dim}--\eqref{eq:derivative-psi} is provided in \ref{sec:appendix-directional-derivative}.
The expressions in \eqref{eq:derivative-psi} are in explicit integral forms and can be evaluated efficiently similar to solution representations \eqref{eq:integral-representation-final-neutron} and \eqref{eq:integral-representation-final-solute}.

\subsection*{Step 3: Discretization of the necessary conditions}
\label{sec:computations-kkt}
In the final step, we show how to solve \eqref{eq:kkt-infinite-dim} by discretizing the variables $u$ and $\lambda_j$.
For brevity, we consider the case where $\lambda_j$ depends only on $t$ for $j=1,2$, i.e., the state constraints in \eqref{eq:optimal-control-reformulate} are only added at a single $x$.
Extension to state constraints added on finite interval $L_1\leq x\leq L_2$ is straightforward and can be regarded as adding multiple state constraints at different $x$ after space discretization.

Let $\pmb{\tau} = \begin{bmatrix}
\tau_0 & \tau_1 & \cdots & \tau_N
\end{bmatrix}$ where $\tau_m = m\Delta t$ for $m=0,\ldots,N$ and $\Delta t = T/N$.
Let $\pmb{u}=\begin{bmatrix}
u(\tau_0) & u(\tau_1) & \cdots & u(\tau_N)
\end{bmatrix}$ denote the discretized control $u$ and $\pmb{\lambda}_j= \begin{bmatrix}
\lambda_j(\tau_0) & \lambda_j(\tau_1) & \cdots & \lambda_j(\tau_N)
\end{bmatrix}$ denote the discretized Lagrange multipliers $\lambda_j,j=1,\ldots,4$. 
Each entry $u(\tau_m)$ and $\lambda_j(\tau_m)$ represents the corresponding values at $t=\tau_m,m=0,\ldots,N$.
Let $u_{\text{pw}}(t)$ be a piecewise constant function defined as
 $u_{\text{pw}}(t) := u(\tau_m)$ if $\tau_m\leq t<\tau_{m+1}$.
Then, the first equation in \eqref{eq:kkt-infinite-dim} can be written in its discretized form:
\begin{equation}\label{eq:derivative-discretized}
    \begin{split}
        &\frac{\delta J}{\delta u}(u_{\text{pw}},\tau_m) + \sum_{j=1}^{4}\frac{\delta G_j}{\delta u}(u_{\text{pw}},\tau_m) = 0, \quad m=0,\ldots,N, \\
    \end{split}
\end{equation}
where 
\begin{align*}
        \frac{\delta J}{\delta u}(u_{\text{pw}},\tau_m) &= 2w_1\int_{\tau_m}^{T}\left[\int_{0}^{\infty}\psi(x,t,u_{\text{pw}}) \frac{\delta\psi}{\delta u} (x,t,u_{\text{pw}},\tau_m)\ dx\right] dt \\ & \quad + 2w_2u(\tau_m),\quad m=0,\ldots,N, \\
    \frac{\delta G_j}{\delta u}(u_{\text{pw}}, \pmb{\lambda}_j, \tau_m) &= (-1)^j\sum_{s = m}^{N-1}\int_{\tau_s}^{\tau_{s+1}}dt\ \lambda_j(\tau_s)\frac{\delta \psi}{\delta u}(x,t,u_{\text{pw}},\tau_m),\\
    &\hspace{1.3in} m=0,\ldots,N,\quad j = 1,2, \\
    \frac{\delta G_j}{\delta u}(u_{\text{pw}}, \pmb{\lambda}_j, \tau_m) &= (-1)^j \lambda_j(\tau_m),\quad m=0,\ldots,N,\quad j=3,4.
\end{align*}

For the inequality constraints in \eqref{eq:kkt-infinite-constraints}, we introduce auxiliary variables $\pmb{z}^g_j = \begin{bmatrix}
z^g_j(\tau_0) & z^g_j(\tau_1) & \cdots & z^g_j(\tau_N)
\end{bmatrix}$ and $\pmb{z}^\lambda_j = \begin{bmatrix}
z^\lambda_j(\tau_0) & z^\lambda_j(\tau_1) & \cdots & z^\lambda_j(\tau_N)
\end{bmatrix}$ for $j=1,\ldots,4$ to rewrite the inequalities as equalities, i.e., 
\begin{equation}
    \label{eq:inequality-constraint-discretized}
    \begin{split}
        g_j(x,\tau_m,u_{\text{pw}}) + z^g_j(\tau_m)^2 &= 0, \quad m=0,\ldots,N,\quad j=1,\ldots,4, \\
    \lambda_j(\tau_m) - z^\lambda_j(\tau_m)^2 &= 0, \quad m=0,\ldots,N,\quad j=1,\ldots,4.
    \end{split}
\end{equation}
Since $\lambda_jg_j\leq0,j=1,\ldots,4$, the last equation $G_j(u) = \int_{0}^{T}\lambda_j(t)g_j(x,t,u_{\text{pw}}) dt$ $=0$ in \eqref{eq:kkt-infinite-dim} is equivalent to $\lambda_j(t)g_j(x,t,u_{\text{pw}})=0$ for all $0\leq t\leq T,j=1,\ldots,4$, or equivalently, the following,
\begin{equation}
    \label{eq:integral-constraint-discretized}
        z_j^g(\tau_m) z_j^\lambda(\tau_m) = 0,\quad m=0,\ldots,N, \quad j=1,\ldots,4.
\end{equation}
We have obtained a system of equations \eqref{eq:derivative-discretized}--\eqref{eq:integral-constraint-discretized} for unknowns $\pmb{u},\pmb{\lambda} , \pmb{z}^g, \pmb{z}^\lambda$ where $\pmb{\lambda} = [\pmb{\lambda}_1\, \cdots\, \pmb{\lambda}_4],\pmb{z}^g = [\pmb{z}^g_1 \, \cdots \, \pmb{z}^g_4]$, and $\pmb{z}^\lambda = [\pmb{z}^\lambda_1\,  \cdots\, \pmb{z}^\lambda_4]$.
This system of equations can be solved using existing solvers, e.g., \zlcomment{ \texttt{fsolve} in MATLAB where the Levenberg-Marquardt method is one of the available algorithms \citep{MATLAB:2023}.
In the following subsection, we present a general form of the Levenberg-Marquardt method with guaranteed global convergence at a quadratic convergence rate.}
\zlcomment{
\subsection{Levenberg-Marquardt method}
\label{sec:numerical-analysis}
In this subsection, we discuss convergence properties of the Levenberg-Marquardt method for our problem.
We consider a general form of the Levenberg-Marquardt algorithm adapted from \citet[Algorithm 9.1.42]{facchinei2003finite} and applied to \eqref{eq:derivative-discretized}--\eqref{eq:integral-constraint-discretized}. 
First, we reformulate equations \eqref{eq:derivative-discretized}--\eqref{eq:integral-constraint-discretized} as the following optimization problem:
\begin{equation}
    \label{eq:kkt-optimization-reformulation}
    \begin{split}
        \min_{\pmb{y} = [\pmb{u}\ \pmb{\lambda}\ \pmb{z}^g\ \pmb{z}^\lambda]} &\quad F(\pmb{y}) \equiv \frac{1}{2} \pmb{f}(\pmb{y})^\top \pmb{f}(\pmb{y}), \\
    \end{split}
\end{equation}
where $\pmb{f}(\pmb{y})$ is a vector-valued function consisting of the left-hand sides in \eqref{eq:derivative-discretized}--\eqref{eq:integral-constraint-discretized}.
Note that an optimal solution to \eqref{eq:kkt-optimization-reformulation} with zero optimal value is also a solution to \eqref{eq:derivative-discretized}--\eqref{eq:integral-constraint-discretized}.
The function $\pmb{f}(\pmb{y})$ is continuously differentiable since the left-hand sides of \eqref{eq:derivative-discretized}--\eqref{eq:integral-constraint-discretized} are compositions of smooth functions on $\pmb{u}, \pmb{\lambda}, \pmb{z}^g,$ and $\pmb{z}^\lambda$.
Then, the gradient and the Hessian of the objective function $F(\pmb{y})$ in \eqref{eq:kkt-optimization-reformulation} are given by
\begin{equation}
    \begin{split}
            &\nabla F(\pmb{y}) = \pmb{J_f}^\top(\pmb{y}) \pmb{f}(\pmb{y}), \\ 
            &\nabla^2 F(\pmb{y}) = \pmb{J_f}^\top(\pmb{y}) \pmb{J_f}(\pmb{y}) + \sum_{l} f_l(\pmb{y}) \nabla^2 f_l(\pmb{y}),
    \end{split}
\end{equation}
where $\pmb{J_f}(\pmb{y})$ is the Jacobian matrix of $\pmb{f}(\pmb{y})$ and $f_l(\pmb{y})$ is the $l$-th entry of $\pmb{f}(\pmb{y})$.

As an extended Gauss-Newton method, the Levenberg-Marquardt method approximates the Hessian by neglecting the second term in $\nabla^2 F(\pmb{y})$ and adding a regularization term to ensure positive definiteness of the approximated Hessian.
Let $\rho:\mathbb{R}^+\to\mathbb{R}^+$ be a continuous function for the regularization term such that $\rho(z)=0$ if and only if $z=0$.
Following \citet[Algorithm 9.1.42]{facchinei2003finite}, the Levenberg-Marquardt method under consideration is given in Algorithm~\ref{alg:gauss-newton}.

The global convergence result of Algorithm~\ref{alg:gauss-newton} for solving \eqref{eq:derivative-discretized}--\eqref{eq:integral-constraint-discretized} is provided in \citet[Theorem 9.1.43]{facchinei2003finite} and stated as follows.
\begin{enumerate}
    \item Every accumulation point $\tilde{\pmb{y}}$ of the sequence $\{\pmb{y}_k\}$ generated by Algorithm~\ref{alg:gauss-newton} is a stationary point of the problem \eqref{eq:kkt-optimization-reformulation}, i.e., $\nabla F(\tilde{\pmb{y}}) = 0$.
    \item Suppose there exists an accumulation point $\tilde{\pmb{y}}^*$ such that the Jacobian $\pmb{J_f}(\tilde{\pmb{y}}^*)$ is nonsingular. Then, $\{\pmb{y}_k\}$ converges to such an accumulation point $\tilde{\pmb{y}}^*$ starting from any initial point $\pmb{y}_0$.
\end{enumerate}
Note that $\tilde{\pmb{y}}^*$ is a desired zero of $\pmb{f}(\pmb{y})$ and thus a solution to \eqref{eq:derivative-discretized}--\eqref{eq:integral-constraint-discretized}. Moreover, it is important to recognize that the nonsingularity of the Jacobian matrix is assumed only at a single point $\tilde{\pmb{y}}^*$ for global convergence.
In contrast, the classical Gauss-Newton method requires nonsingular Jacobian matrices along the sequence $\{\pmb{y}_k\}$ at every $k$ for convergence \citep[Theorem 10.1]{nocedal2006numerical}.

\begin{remark}
    The key enabler in developing our globally convergent numerical method is that the first-order optimality condition \eqref{eq:kkt-infinite-dim} does not contain differential operators but only integral equations and inequalities. This is due to the reformulation \eqref{eq:optimal-control-reformulate} by replacing the PDE \eqref{eq:general-second-order} with an integral representation of the PDE solution.
    In contrast, existing studies \citep{casas2025bilinear} derive first-order optimality condition as a coupled system of PDEs and a projection operator for the case of simple control constraint, for which only locally convergent numerical method has been developed.
\end{remark}
}
\begin{algorithm}[t]
\caption{\zlcomment{Levenberg-Marquardt method}}
\label{alg:gauss-newton}
\begin{algorithmic}[1]
\zlcomment{
\Require $\gamma\in(0,1)$, initial point $\pmb{y}_0$, direction $\pmb{d}_0$, step size parameter $\beta_0$,
\For{$k=0,1,2,\ldots$}
    \State Compute $\pmb{f}(\pmb{y}_k)$, $\pmb{J_f}(\pmb{y}_k)$; define
    \[
        \psi_k(\pmb{d})=\nabla F(\pmb{y}_k)^\top \pmb{d}+\tfrac{1}{2}\, \pmb{d}^\top \left[\pmb{J_f}^\top(\pmb{y}_k) \pmb{J_f}(\pmb{y}_k) + \rho(F(\pmb{y}_k))\pmb{I}\right]\pmb{d} .
    \]
    \State Compute the direction $\pmb{d}_k= \arg\min_{\pmb{d}} \psi_k(\pmb{d})$.
    \If
    {$ F(\pmb{y}_k + \pmb{d}_k) \leq \gamma F(\pmb{y}_k) $}
    set $\beta_k=0$.
    \Else 
    
    Find the smallest nonnegative integer $\beta_k$ such that
    \[
        F(\pmb{y}_k + 2^{-\beta_k} \pmb{d}_k) \le F(\pmb{y}_k) + \gamma \,2^{-\beta_k} \nabla F(\pmb{y}_k)^\top \pmb{d}_k.
    \]
    \EndIf
    \State Set $\pmb{y}_{k+1} = \pmb{y}_k + 2^{-\beta_k} \pmb{d}_k$.
\EndFor
}
\end{algorithmic}
\end{algorithm}
\zlcomment{
    In addition to global convergence, Algorithm~\ref{alg:gauss-newton} also achieves a quadratic convergence rate that is consistent with \cite{casas2025bilinear}. If $\rho(F(\pmb{y})) = \mathcal{O}(F(\pmb{y}))$ for all $\pmb{y}$ sufficiently near $\tilde{\pmb{y}}^*$ where $\mathcal{O}$ is the asymptotic notation, then the convergence rate is Q-quadratic, i.e., $\limsup_{k\to\infty} \|\pmb{y}_{k+1} - \pmb{y}_\infty\|/\|\pmb{y}_k - \pmb{y}_\infty\|^2 <\infty$.
    Unlike existing studies that require the initial point to be sufficiently close to an optimal solution to achieve quadratic convergence \citep{casas2025bilinear}, Algorithm~\ref{alg:gauss-newton} achieves quadratic convergence starting from an arbitrary initial point under the above condition on $\rho$.
\begin{remark}
    Algorithm~\ref{alg:gauss-newton} does not specify the explicit form of the function $\rho$ in the algorithm. The choice of $\rho$ is a delicate matter in practice \citep[pg.701]{facchinei2003finite}. In computational experiments, we rely on \texttt{fsolve} that uses its own built-in function for $\rho$.
\end{remark}
\begin{remark}
    In Step~\ref{sec:computations-kkt} of our computational framework, we discretize and reformulate the KKT conditions \eqref{eq:kkt-infinite-dim} as a system of smooth nonlinear equations \eqref{eq:derivative-discretized}--\eqref{eq:integral-constraint-discretized} with auxiliary variables $\pmb{z}^g$ and $\pmb{z}^\lambda$. An alternative way is to reformulate the KKT conditions as a system of nonsmooth equations, e.g., by replacing the inequalities and the complementarity condition $G_j(u^*,\lambda^*)=0$ in \eqref{eq:kkt-infinite-dim} with a square-root or point-wise minimum function \citep[Section 1.5]{facchinei2003finite}.
    The nonsmooth reformulation does not introduce auxiliary variables, and the resulting nonsmooth equation can be solved with the same convergence guarantee as Algorithm~\ref{alg:gauss-newton} for \eqref{eq:kkt-optimization-reformulation} by a nonsmooth version of the Levenberg-Marquardt method \citep[Algorithm 9.1.42]{facchinei2003finite}. Despite the presence of auxiliary variables, we found that \texttt{fsolve} is efficient for solving \eqref{eq:derivative-discretized}--\eqref{eq:integral-constraint-discretized} in our computational experiments and we did not pursue the nonsmooth method.
\end{remark}
}

\begin{figure}
\centering
    \begin{subfigure}{0.49\textwidth}
            \includegraphics[width=0.95\textwidth]{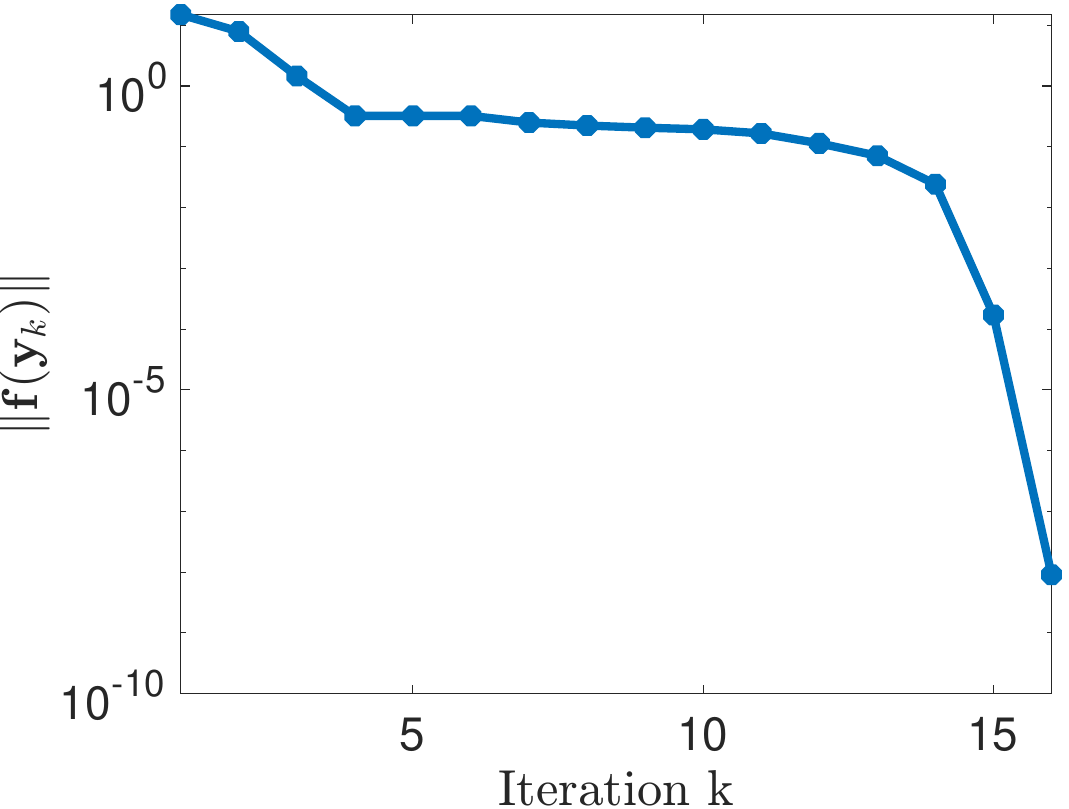}
            \caption{Nuclear reactivity control: convergence}
            \label{fig:convergence-nuclear-norm}
        \end{subfigure}%
        \begin{subfigure}{0.49\textwidth}
            \includegraphics[width=0.95\textwidth]{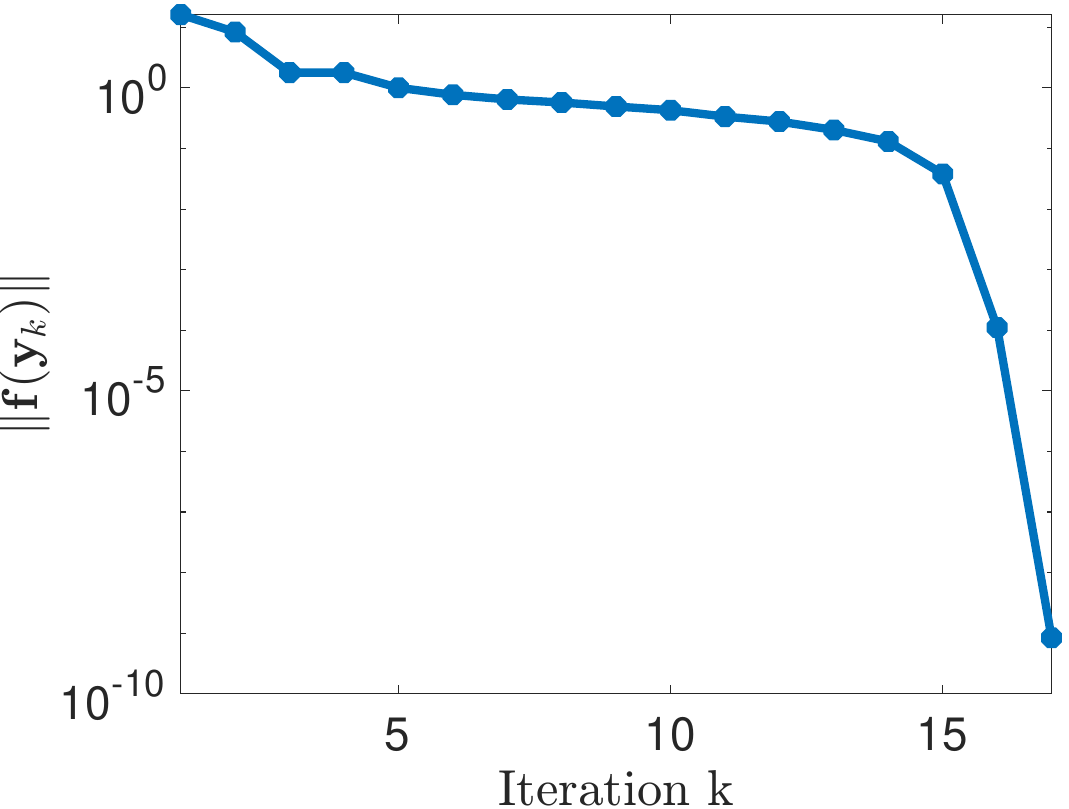}
            \caption{Solute transport in fluids: convergence  }
            \label{fig:convergence-solute-norm}
        \end{subfigure}
        \begin{subfigure}{0.49\textwidth}
            \includegraphics[width=0.95\textwidth]{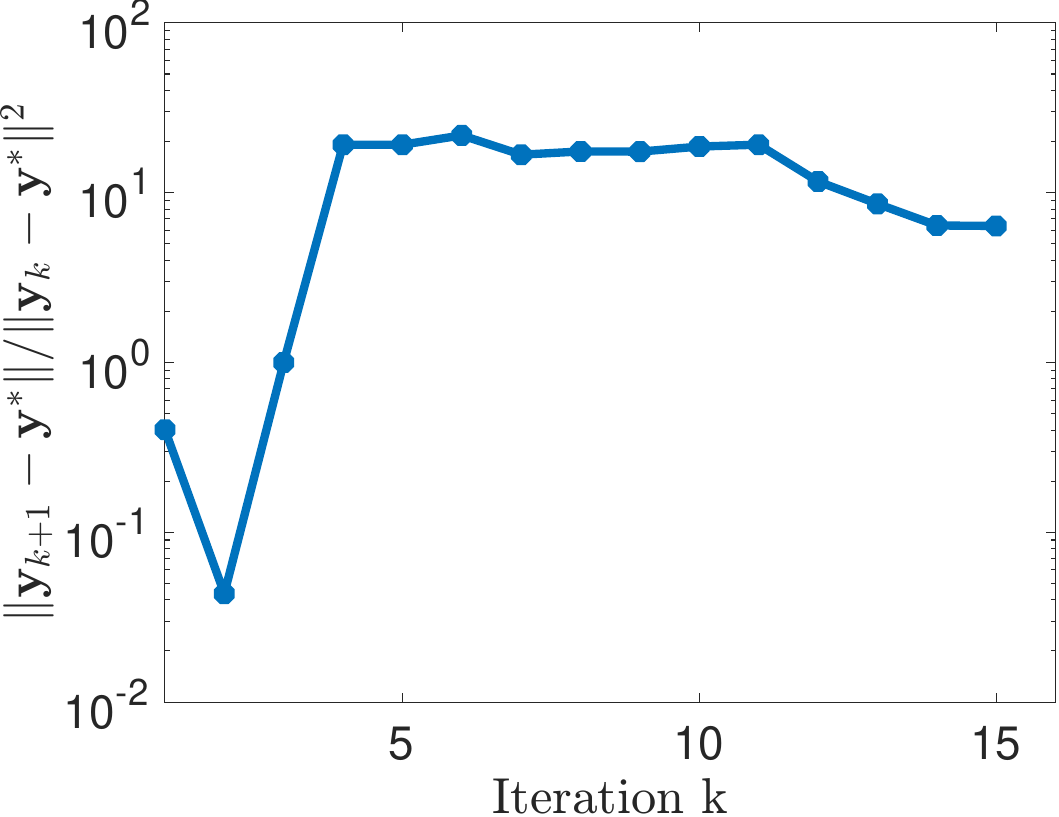}
            \caption{Nuclear reactivity control: Q-quadratic rate}
            \label{fig:convergence-nuclear}
        \end{subfigure}%
        \begin{subfigure}{0.49\textwidth}
            \includegraphics[width=0.95\textwidth]{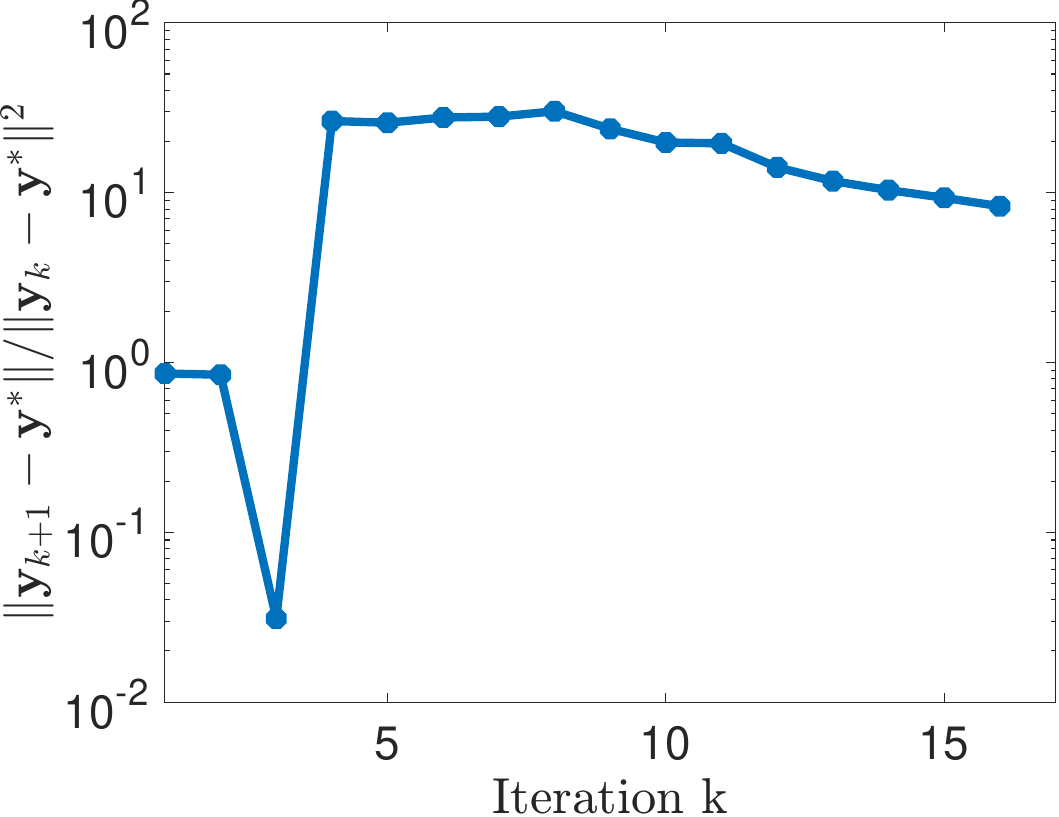}
            \caption{Solute transport in fluids: Q-quadratic rate}
            \label{fig:convergence-nuclear}
        \end{subfigure}
    \caption{\zlcomment{Illustration of Q-quadratic convergence of \texttt{fsolve} using the Levenberg-Marquardt method for two application problems: nuclear reactivity control in Section~\ref{sec:nuclear-reactor} and solute transport in fluids in Section~\ref{sec:ADE}. Following the notation used in Section~\ref{sec:numerical-analysis}, (a) and (b) show that the sequence generated by \texttt{fsolve} converges to a zero of $\pmb{f}(\pmb{y}_k)$; (c) and (d) show that the convergence rate is Q-quadratic since $\|\pmb{y}_{k+1} - \pmb{y}^*\| / \|\pmb{y}_{k} - \pmb{y}^*\|^2$ remains bounded, where $\pmb{y}^*$ is the converged solution.}}
    \label{fig:convergence}
\end{figure}
\section{Computational results}
\label{sec:results}
In this section, we apply the computational framework developed in Section~\ref{sec:computation-framework} to the two application problems in Sections~\ref{sec:nuclear-reactor} and \ref{sec:ADE}.
All results are reported in dimensionless forms.
For each application, we first numerically verify the integral representations \eqref{eq:integral-representation-final-neutron} and \eqref{eq:integral-representation-final-solute} against fully numerical solutions.
We choose MATLAB's built-in function \texttt{pdepe} as a benchmark for PDE solutions.
This function runs a variable time-stepping method and a finite-element method that is second-order accurate in space \citep{skeel1990method}. 
The discretization size is chosen as $\Delta x = 0.01$ and $\Delta t = 0.01$ for space and time, respectively.

Subsequently, we compute optimal control using discretized KKT conditions \eqref{eq:derivative-discretized}--\eqref{eq:integral-constraint-discretized}.
All numerical computations were performed in MATLAB.
For the computation of integral representations \eqref{eq:integral-representation-final-neutron} and \eqref{eq:integral-representation-final-solute}, the MATLAB function \texttt{integral} \citep{shampine2008vectorized} is used for the line integral along $\partial\mathcal{D}^+$, where $\partial\mathcal{D}^+$ is the union of the two line segments
\begin{align*}
    &\{r\left[\cos(\pi-\theta) + i\sin(\pi - \theta)\right]:r\geq0\}, \quad\{r\left[\cos(\theta) + i\sin(\theta)\right]:r\geq0\},
\end{align*}
with $\theta=\pi/8$.
The optimality conditions \eqref{eq:derivative-discretized}--\eqref{eq:integral-constraint-discretized} are solved with the MATLAB function \texttt{fsolve} using the option of Levenberg-Marquardt method. 
The convergence tolerance is set to $10^{-6}$, and the solver converges in less than 30 iterations for all numerical experiments.
All computations were performed on a laptop with an Apple M2 chip and 8 GB RAM.
\zlcomment{Example runs for the two applications specified in Sections~\ref{sec:nuclear-reactor} and \ref{sec:ADE} are reported in Figure~\ref{fig:convergence} to illustrate Q-quadratic convergence.
To test whether the converged solution to first-order optimality conditions \eqref{eq:derivative-discretized}--\eqref{eq:integral-constraint-discretized} is a local minimum, for each case we used ten randomized initial points in \texttt{fsolve} and found that they all converge to the same solution.
}

\subsection{Nuclear reactivity control}

\begin{figure}[t]
    \begin{subfigure}{0.49\textwidth}
            \includegraphics[width=0.95\textwidth]{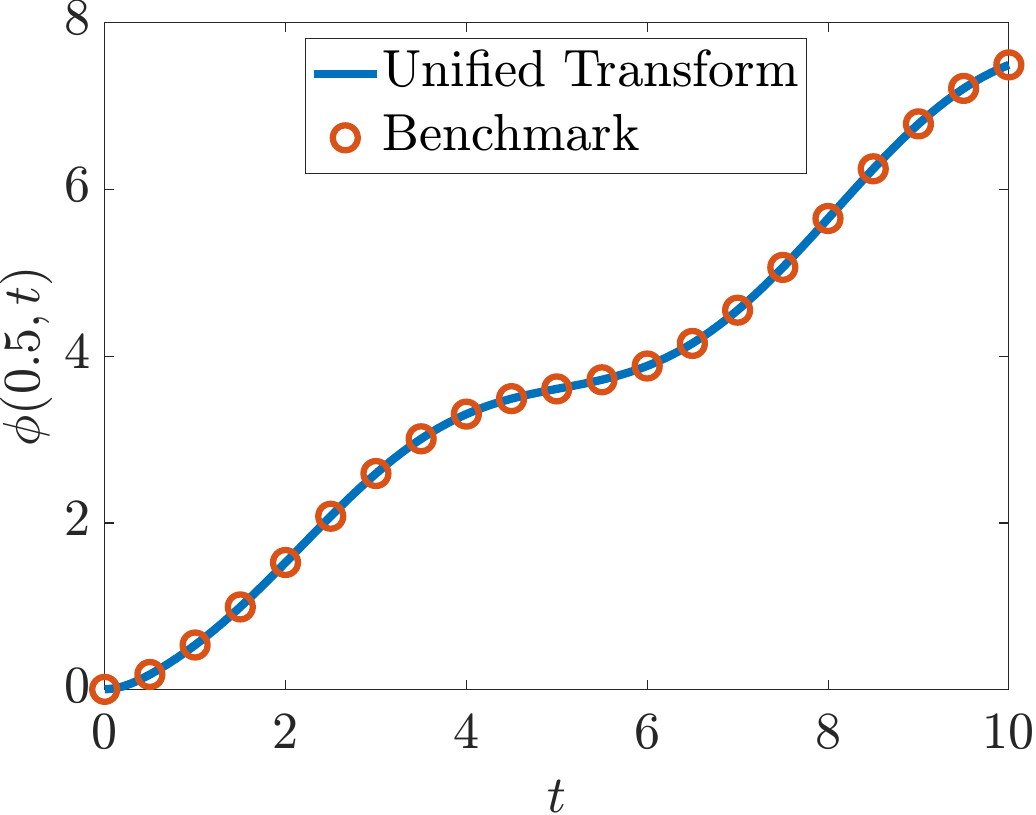}
            \caption{Neutron flux at $x=0.5$ }
            \label{fig:solution-compare-nuclear-x-1}
        \end{subfigure}%
        \begin{subfigure}{0.49\textwidth}
            \includegraphics[width=0.95\textwidth]{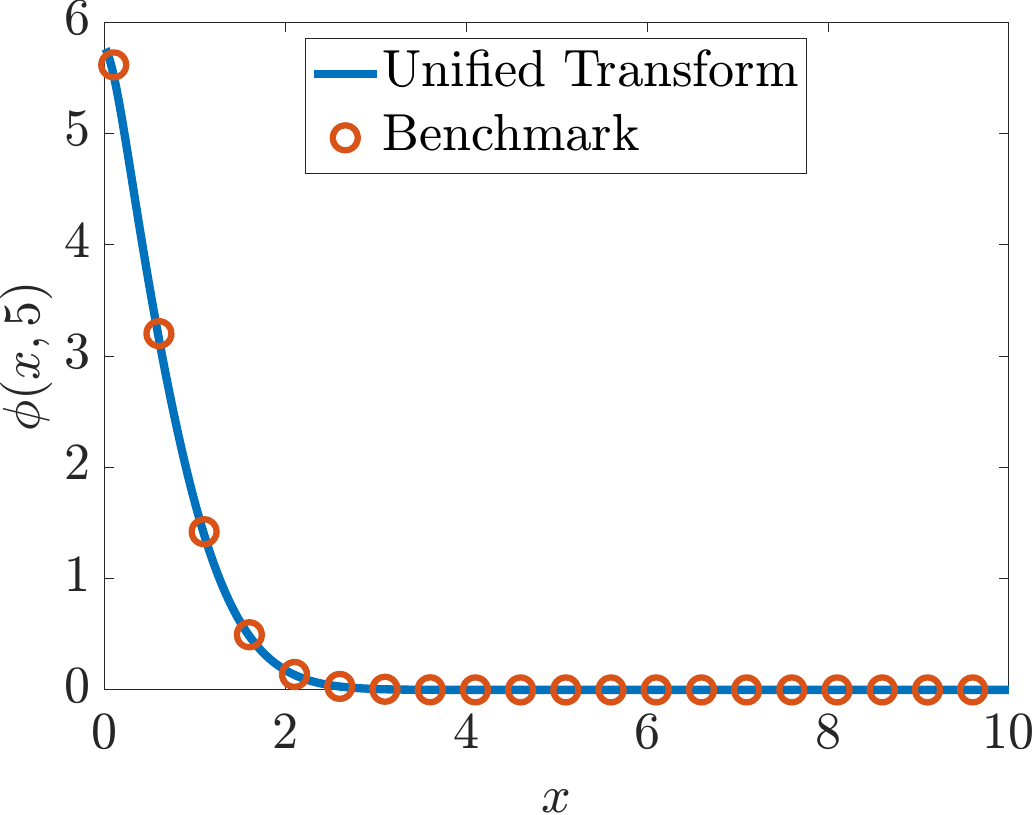}
            \caption{Neutron flux at $t=5$  }
            \label{fig:solution-compare-nuclear-t}
        \end{subfigure}
    \caption{Comparison between the semi-analytical solution derived from the unified transform method (see \eqref{eq:integral-representation-final-neutron}) with a fully numerical solution. 
    Dimensionless neutron flux versus dimensionless time (left) and versus dimensionless distance (right).}
    \label{fig:comparison-nuclear}
    \end{figure}
\subsubsection{Verification of the semi-analytical solution}
Following the problem in Section~\ref{sec:nuclear-reactor}, we first numerically verify the semi-analytical solution \eqref{eq:dimensionless-diffusion-equation-neutron} to \eqref{eq:integral-representation-final-neutron} using the following initial and boundary conditions:
\begin{align*}
    & \phi_o(x) = e^{-10x}, \quad\quad \xi_{\text{Ne}}(t) =-10(1 + \sin(t) / 2). \label{eq:example-comparison-0}
\end{align*}

Following the parameters reported in the literature \citep[pg.$\,$211]{duderstadt1976nuclear}, we set $D_n=9.21$ (cm) and $\bar{\Sigma}_a = 0.1532$ (cm$^{-1}$). 
We consider a time-varying absorption coefficient $\Sigma_a(t) = \bar{\Sigma}_a e^{-t}$.
Figure~\ref{fig:comparison-nuclear} shows that the numerical evaluation of \eqref{eq:integral-representation-final-neutron} derived from the unified transform method agrees with the benchmark solution computed by MATLAB \texttt{pdepe}.

\begin{figure}[tb]
    \begin{subfigure}{0.49\textwidth}
        \centering
        \includegraphics[width=0.95\textwidth]{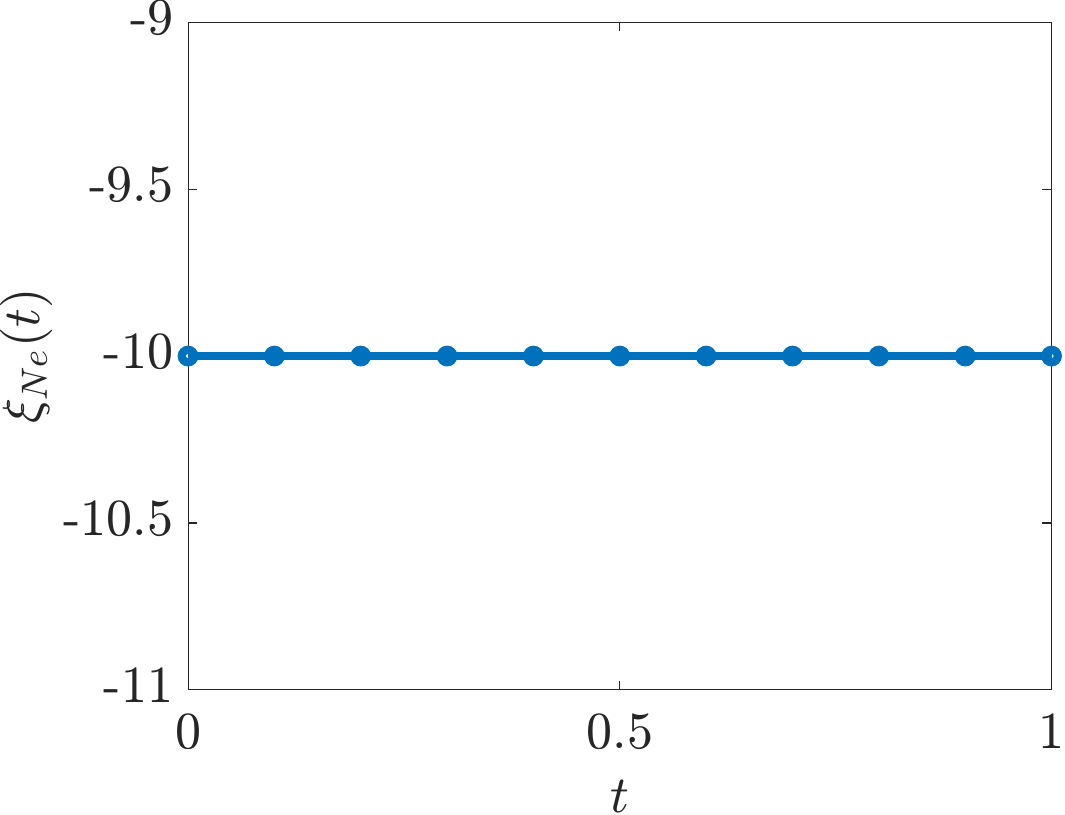}
        \caption{Boundary condition $\xi_{\text{Ne}}(t)$}
        \label{fig:boundary-condition-nuclear-constant}    
    \end{subfigure}\hfill
    \begin{subfigure}{0.49\textwidth}
        \centering
        \includegraphics[width=0.95\textwidth]{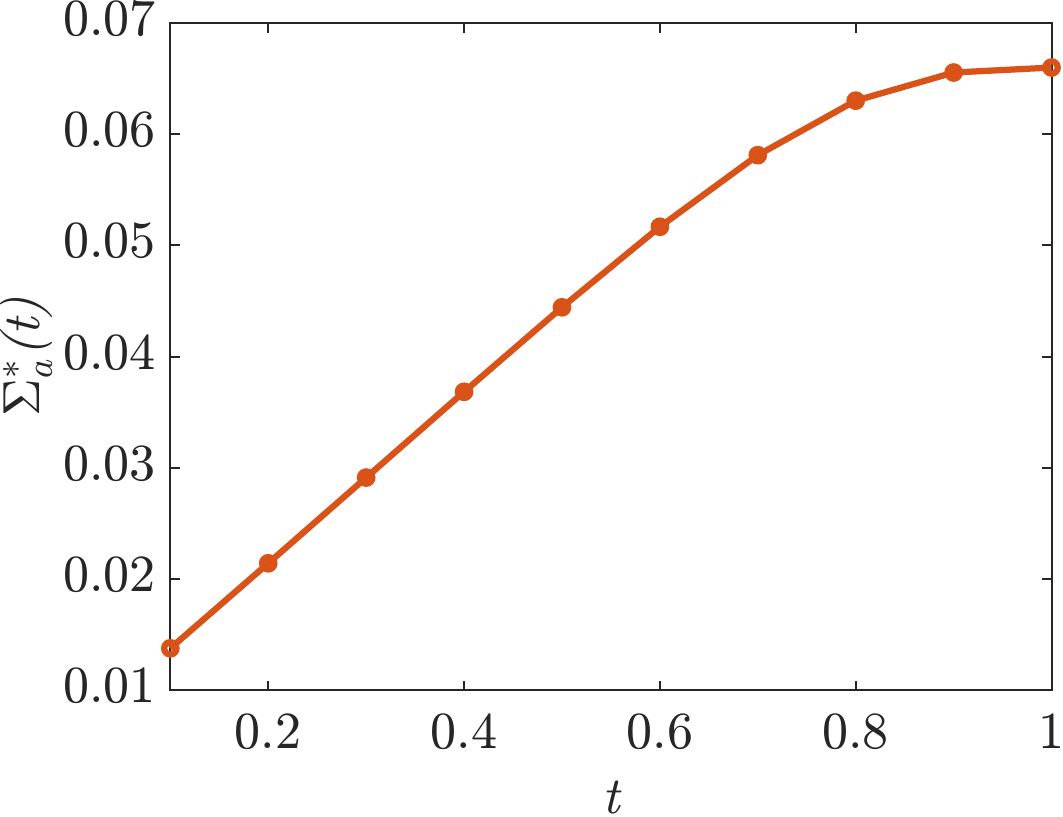}
        \caption{Optimal control $\Sigma^*_a(t)$}
        \label{fig:optimal-rate-nuclear-constant}
    \end{subfigure}
    \begin{subfigure}{0.49\textwidth}
        \centering
        \includegraphics[width=\textwidth]{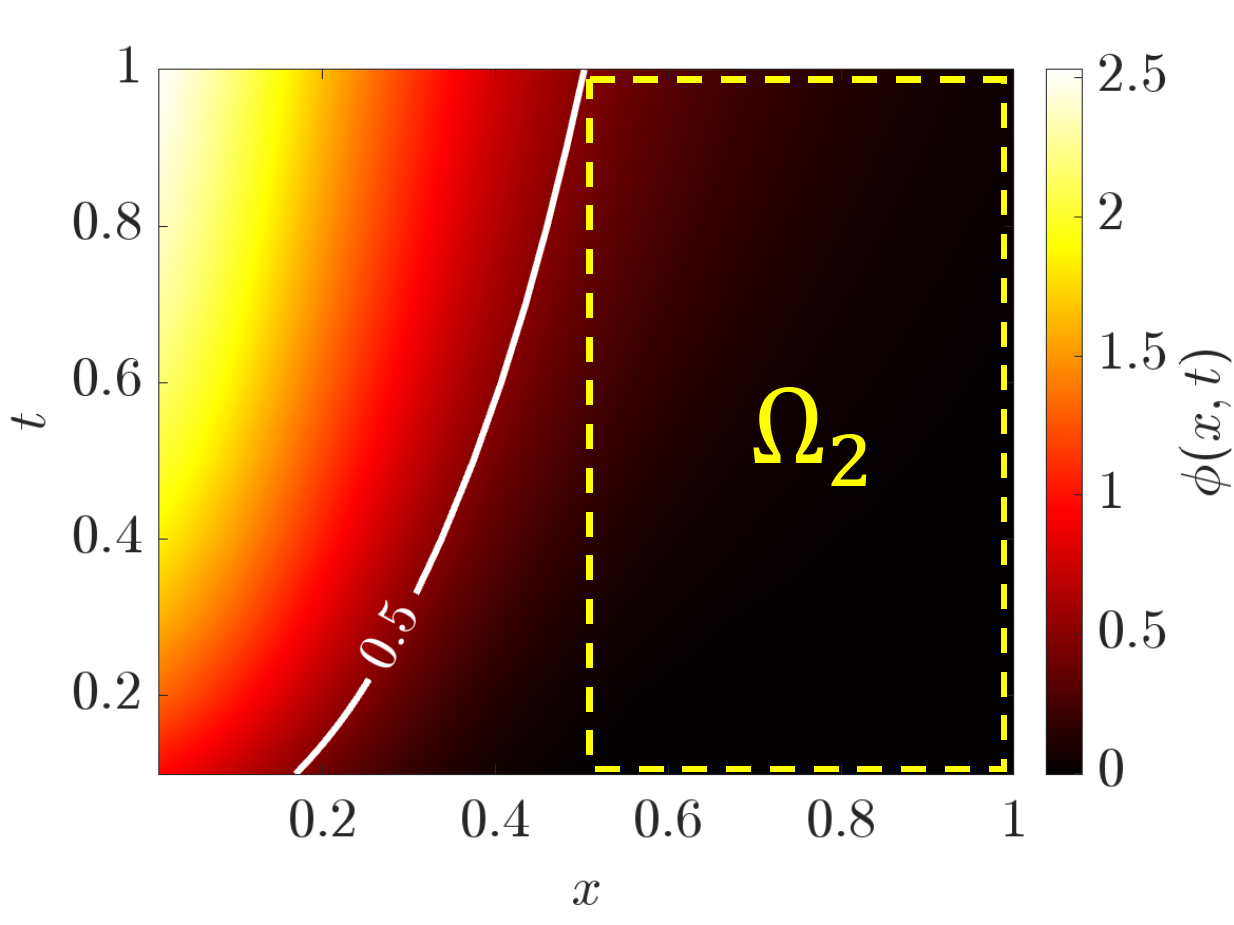}
        \caption{Neutron flux $\phi(x,t)$}
        \label{fig:state-contour-nuclear-constant}
    \end{subfigure}\hfill  
    \begin{subfigure}{0.49\textwidth}
    \centering
        \includegraphics[width=0.95\textwidth]{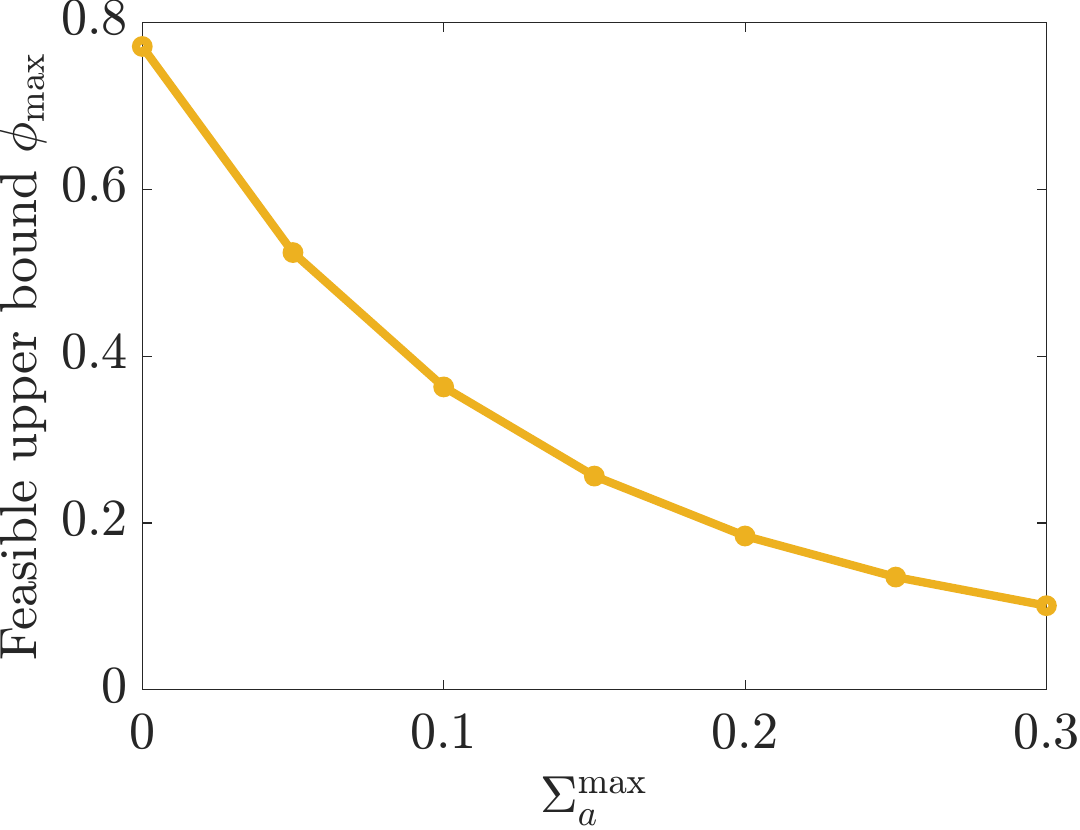}
        \caption{Feasible upper bound $\phi_{\max}$}
        \label{fig:maximal-Upper-bound} 
    \end{subfigure}
    \caption{Nuclear reactivity control with constant boundary condition specified in (a).
    The computed optimal control $\Sigma^*_a(t)$ is shown in (b).
    The neutron flux $\phi(x,t)$ under optimal control is shown in (c) where $\phi(x,t)\leq 0.5$ is imposed in the region $\Omega_2 = \{x\geq 0.5, 0\leq t\leq 1\}.$
    (d) gives minimal $\phi_{\max}$ that is feasible for \eqref{eq:optimal-control-neutron} for different values of $\Sigma_a^{\max}$.
    }
    \label{fig:nuclear-reactivity-control-constant}
\end{figure}
\begin{figure}
    \begin{subfigure}{0.49\textwidth}
        \centering
        \includegraphics[width=0.95\textwidth]{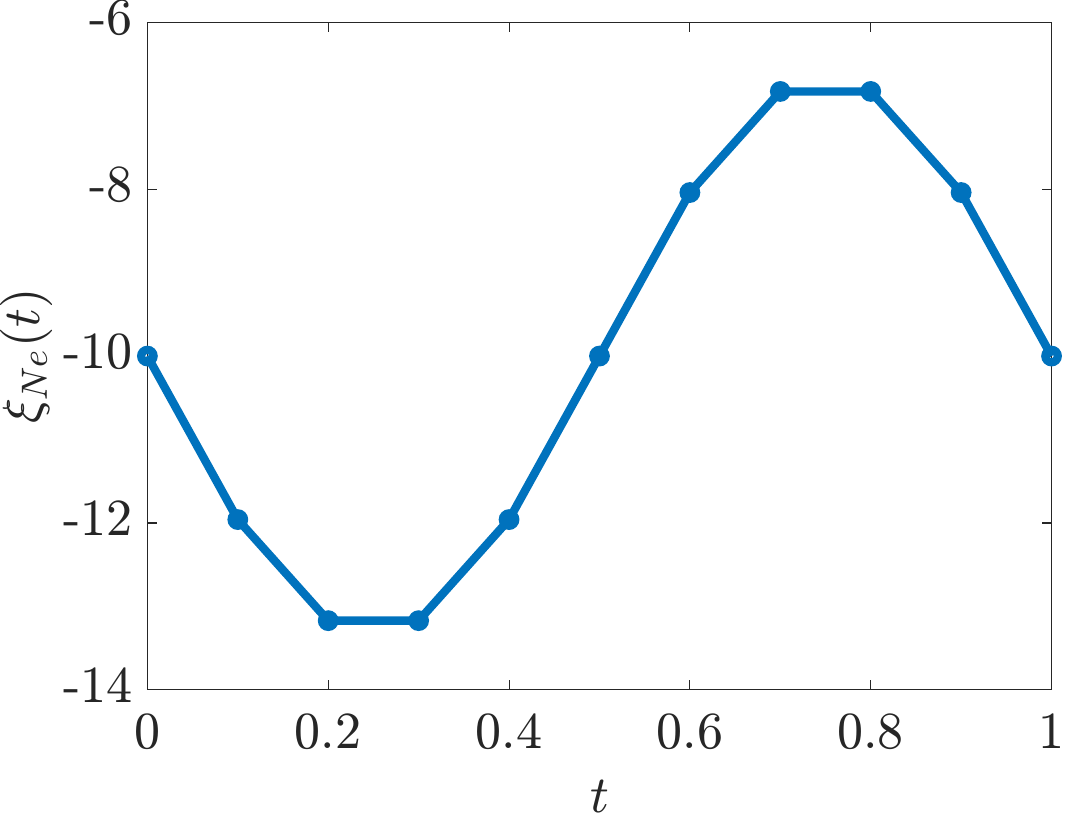}
        \caption{Boundary condition $\xi_{\text{Ne}}(t)$}
        \label{fig:boundary-condition-nuclear-sin}    
    \end{subfigure}\hfill
    \begin{subfigure}{0.49\textwidth}
        \centering
        \includegraphics[width=0.95\textwidth]{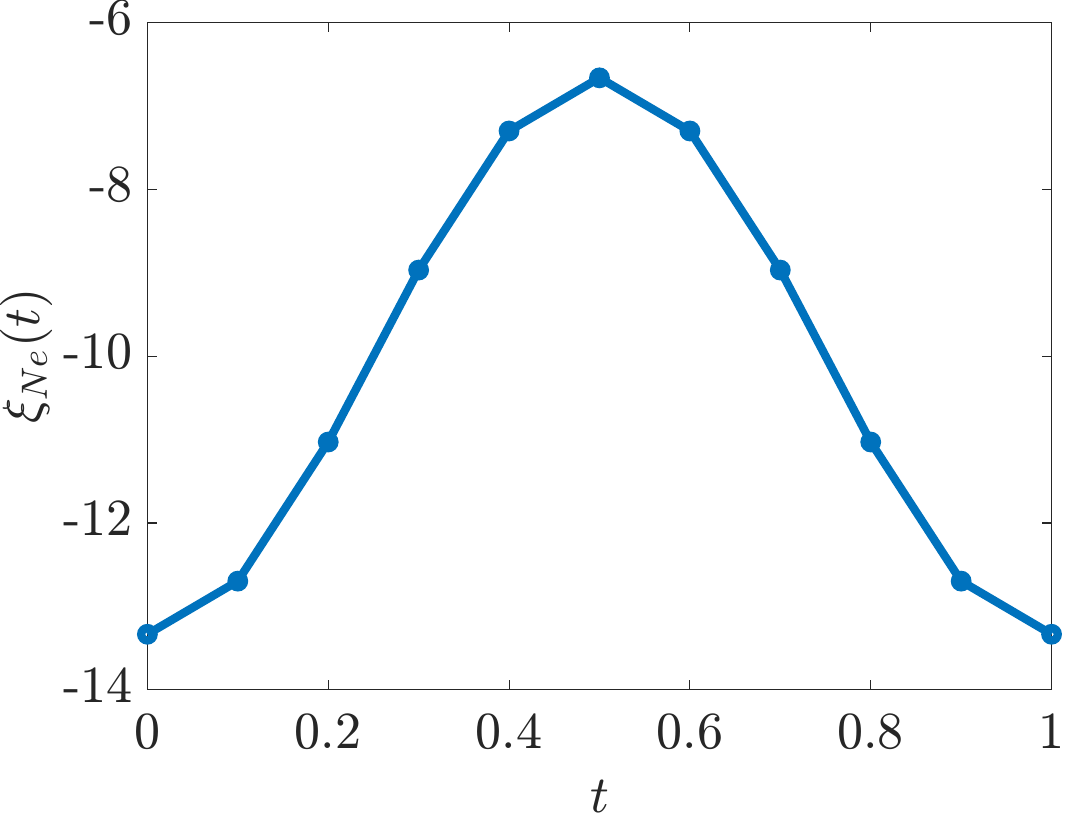}
        \caption{Boundary condition $\xi_{\text{Ne}}(t)$}
        \label{fig:boundary-condition-nuclear-cos}
    \end{subfigure}\hfill
    \begin{subfigure}{0.49\textwidth}
        \centering
        \includegraphics[width=0.95\textwidth]{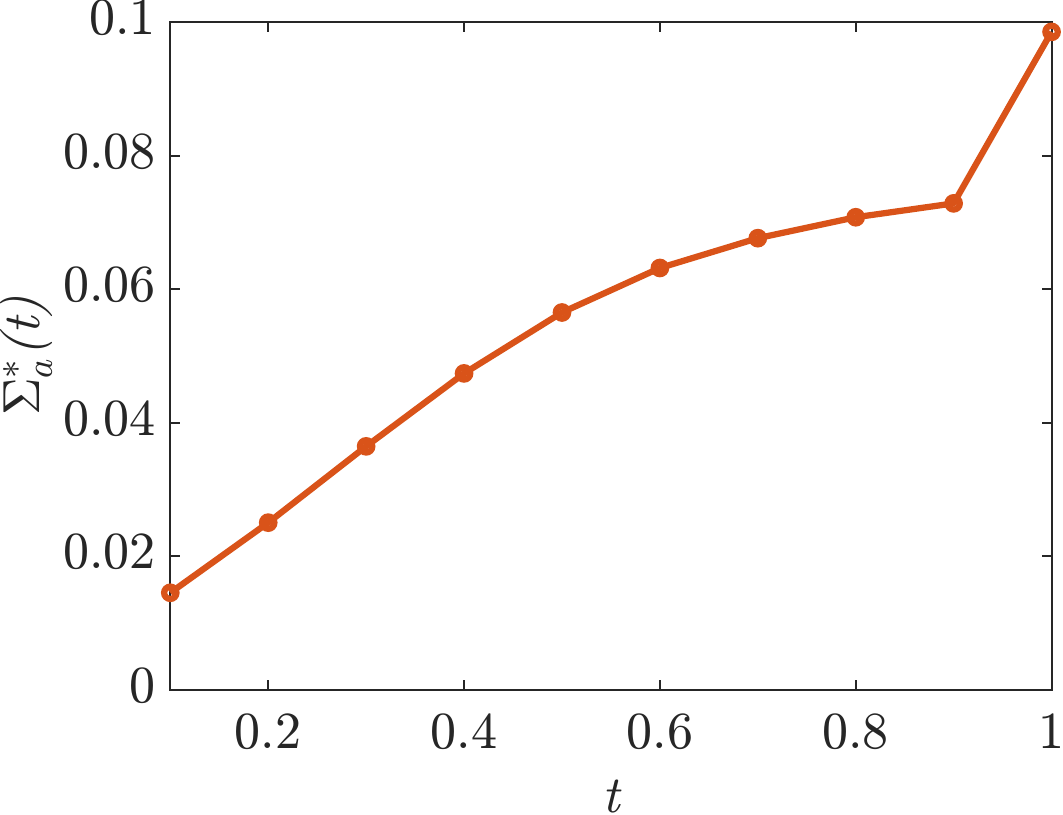}
        \caption{Optimal control $\Sigma_a^*(t)$}
        \label{fig:optimal-control-nuclear-cos}    
    \end{subfigure}\hfill
    \begin{subfigure}{0.49\textwidth}
        \centering
        \includegraphics[width=0.95\textwidth]{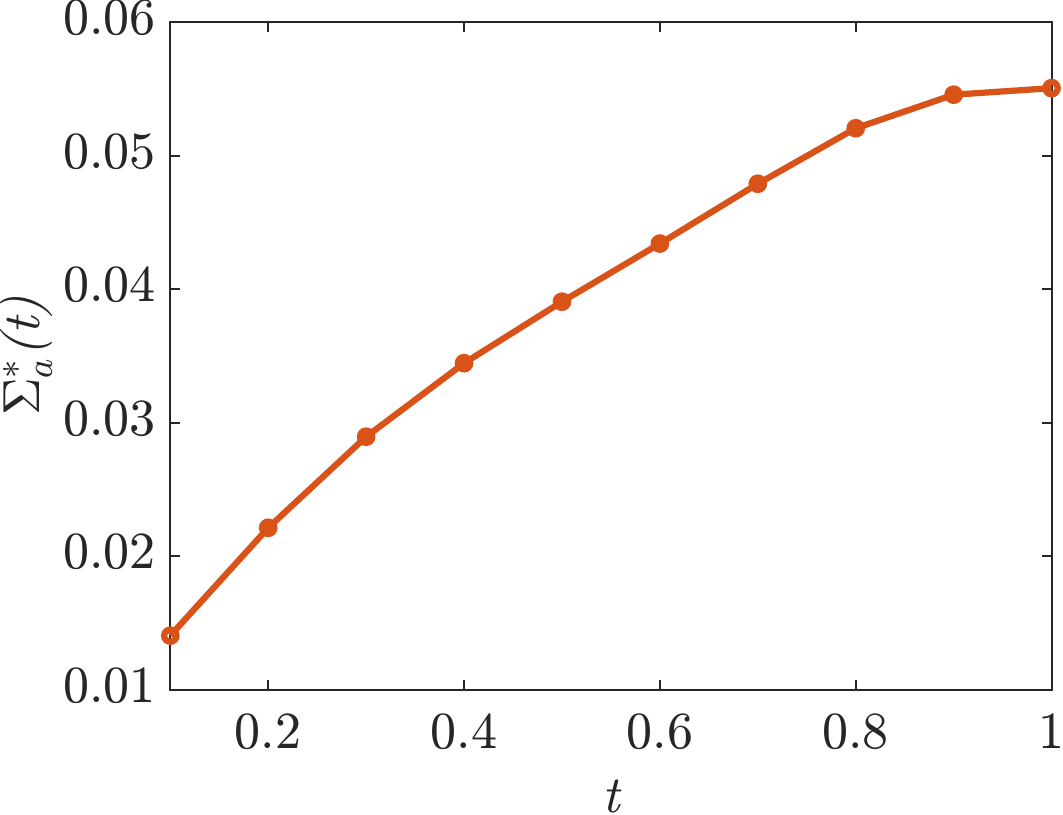}
        \caption{Optimal control $\Sigma_a^*(t)$}
        \label{fig:optimal-control-nuclear-sin}
    \end{subfigure}
    \begin{subfigure}{0.49\textwidth}
        \centering
        \includegraphics[width=\textwidth]{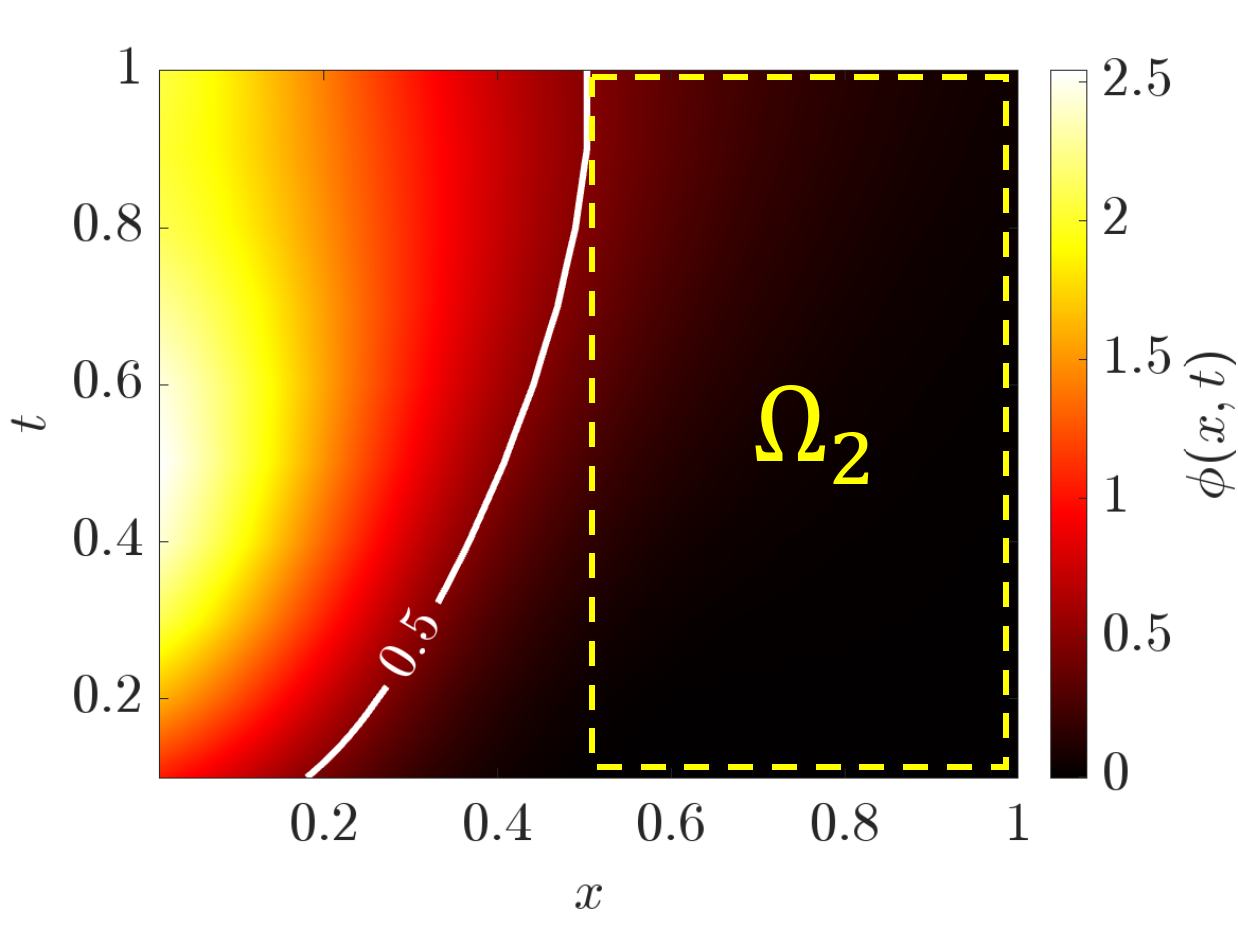}
        \caption{Neutron flux $\phi(x,t)$}
        \label{fig:state-contour-nuclear-sin}
    \end{subfigure}
    \begin{subfigure}{0.49\textwidth}
        \centering
        \includegraphics[width=\textwidth]{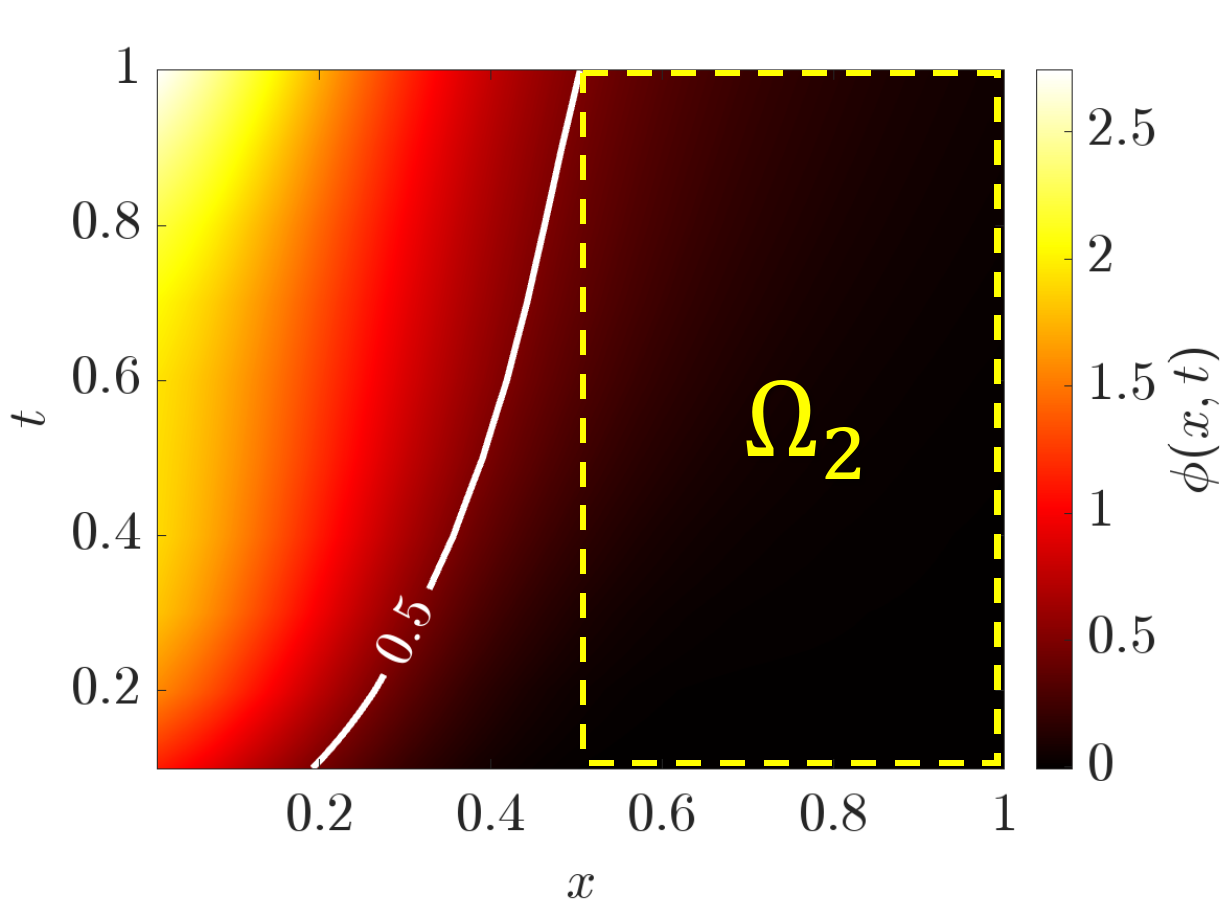}
        \caption{Neutron flux $\phi(x,t)$}
        \label{fig:state-contour-nuclear-cos}
    \end{subfigure}
    \caption{Nuclear reactivity control with time-varying boundary conditions specified in (a) and (b), respectively. The computed optimal control is shown in (c) and (d), respectively. The neutron flux under the computed control is shown in (e) and (f), respectively. The state constraint $\phi(x,t)\leq 0.5$ is imposed in the region $\Omega_2 = \{x\geq 0.5, 0\leq t\leq 1\}.$}
    \label{fig:nuclear-reactivity-control-time-varying}
\end{figure}

\subsubsection{Optimal control}
The main goal of adding chemical shim to the reactor core is to reduce the neutron flux below a desired level to reduce the reactivity of the reactor core.
It is also desirable to minimize the amount of chemical shim such as boron added to the reactor core \citep{do2020physics}.
Therefore, we consider the following optimal control problem to achieve these goals:
\begin{equation}\label{eq:optimal-control-neutron}
    \begin{split}
        &\min_{0\leq \Sigma_a\leq\Sigma^{\max}_a} \quad \int_{0}^{T} \Sigma_a^2(t) dt\\
        & \text{s.t.}\quad \phi(x,t,\Sigma_a)\leq \phi_{\max},\quad  (x,t)\in\Omega_2=\{x\geq L_1,0\leq t\leq T\},\\
    \end{split}
\end{equation}
where $T = 1$, $L_1=0.5, \phi_{\max} = 0.5$.
This represents the case where we want to keep the neutron flux $\phi$ below $0.5$ in the region $\Omega_2$.
The initial condition is set to $\phi_o(x)=e^{-10x}$.
Figure~\ref{fig:nuclear-reactivity-control-constant} shows the results with constant boundary condition $\xi_{\text{Ne}}(t)=-10$.
The constant boundary condition corresponds to the case where neutrons are emitted from the source at a constant rate.

As shown in Figure~\ref{fig:optimal-rate-nuclear-constant}, 
optimal control $\Sigma_a^*(t)$ increases monotonically as $t$ increases.
The upper bound $\phi_{\max}=0.5$ is active \zlcomment{only when $t=1$, meaning that the state value is strictly lower than $\phi_{\max}$ for $t<1$ in the entire region $\Omega_2$,}  as shown in Figure~\ref{fig:state-contour-nuclear-constant}. 
\zlcomment{In practice}, increasing the absorption cross-section $\Sigma_a$ corresponds to adding more chemical shim to the reactor core, and thus reducing the neutron flux value $\phi$. 
\zlcomment{The optimal strategy for controlling $\Sigma_a$ in Figure~\ref{fig:optimal-rate-nuclear-constant}} suggests that the upper bound $\phi_{\max}$ is reached only when $t=1$ at $x=0.5$.
\zlcomment{The orange curve in Figure~\ref{fig:maximal-Upper-bound} shows the minimum values of the upper bound $\phi_{\max}$ for \eqref{eq:optimal-control-neutron} to be feasible under different values of $\Sigma_a^{\max}$.}
In other words, the upper bound $\phi_{\max}$ is required to be above the orange curve in Figure~\ref{fig:maximal-Upper-bound}.
This reflects the different requirements we can impose on the neutron flux based on different amounts of chemical shim allowed in the reactor core.

Figure~\ref{fig:nuclear-reactivity-control-time-varying} illustrates the results under different time-varying boundary conditions.
For the purpose of illustration, the boundary conditions in Figures~\ref{fig:boundary-condition-nuclear-sin}--\ref{fig:boundary-condition-nuclear-cos} are set to $\xi_{\text{Ne}}(t)=-10(1+\sin(2\pi t)/2)$ and $\xi_{\text{Ne}}(t)=-10(1+\cos(2\pi t)/2)$, respectively, with the same average value $-10$ as in Figure~\ref{fig:boundary-condition-nuclear-constant}.
Comparing Figure~\ref{fig:optimal-rate-nuclear-constant} with Figures~\ref{fig:optimal-control-nuclear-cos} and \ref{fig:optimal-control-nuclear-sin}, optimal control $\Sigma_a^*(t)$ exhibits similar trends, but \zlcomment{the peak values of $\Sigma_a^*(t)$ vary under} different types of boundary conditions.
Specifically, \zlcomment{optimal control
$\Sigma_a^*(t)$ for $t=1$ in Figure~\ref{fig:optimal-control-nuclear-cos} is larger than the peak value in Figure~\ref{fig:optimal-rate-nuclear-constant}, and the latter is larger than the peak value in Figure~\ref{fig:optimal-control-nuclear-sin}.
Therefore, the optimal amount of chemical shim depends not only on the average emission rate of neutrons but also on how neutrons are emitted at the boundary in terms of amplitude and frequency.}

\subsection{Solute transport in fluids}
\subsubsection{Verification of the semi-analytical solution}
\begin{figure}[t]
    \begin{subfigure}{0.49\textwidth}
            \includegraphics[width=0.95\textwidth]{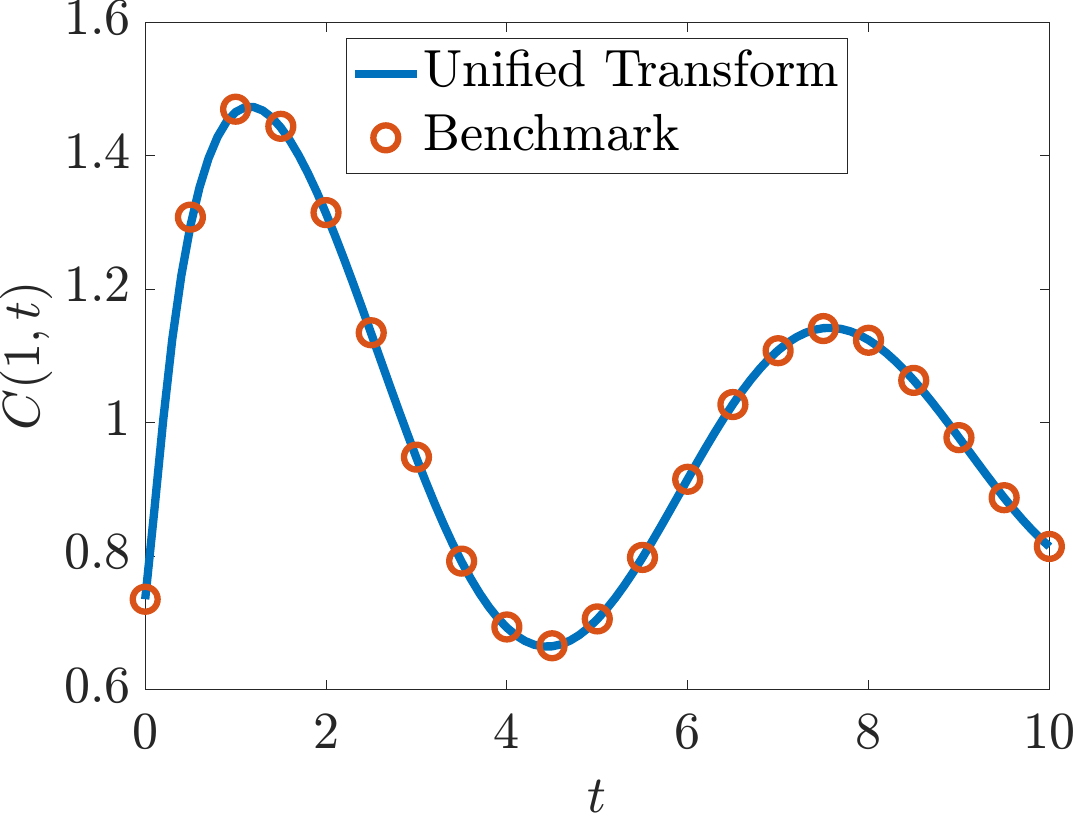}
            \caption{Solute concentration at $x=1$ }
            \label{fig:comparison-constant-time-ic-0}
        \end{subfigure}%
        \begin{subfigure}{0.49\textwidth}
            \includegraphics[width=0.95\textwidth]{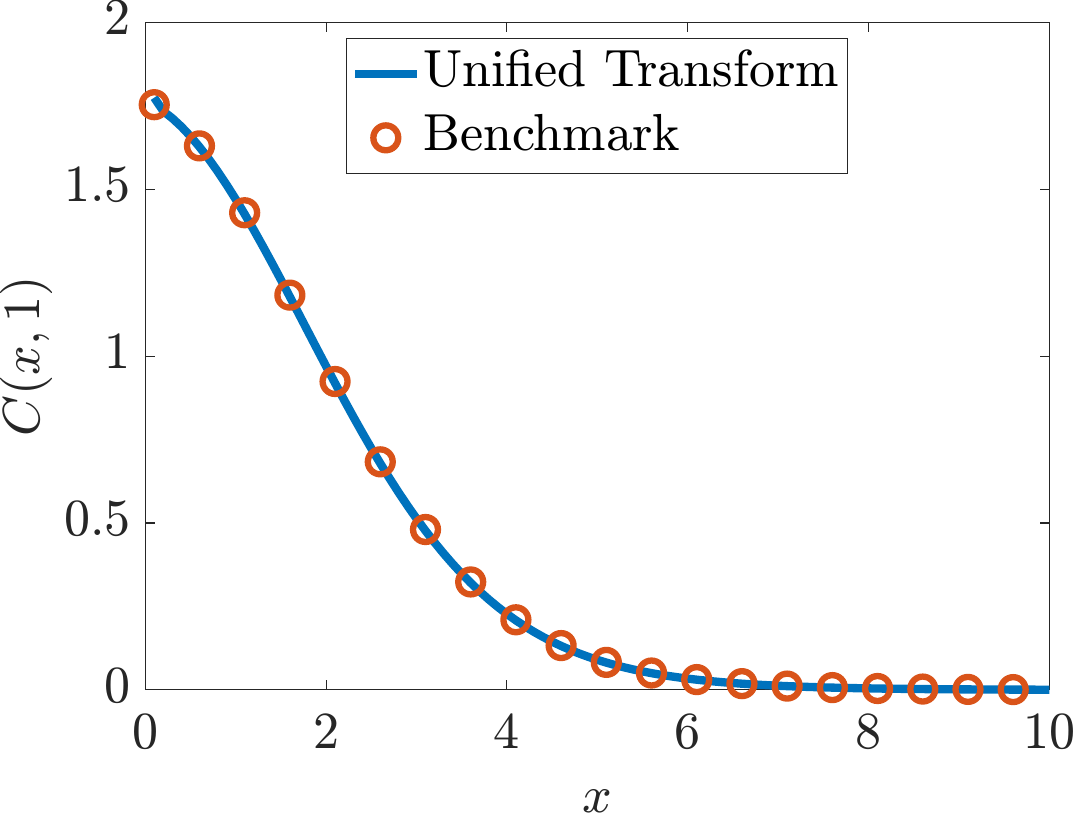}
            \caption{Solute concentration at $t=1$  }
            \label{fig:comparison-constant-space-ic-0}
        \end{subfigure}
    \caption{Comparison between the semi-analytical solution derived from the unified transform method (see \eqref{eq:integral-representation-final-solute}) with a fully numerical solution. 
    Dimensionless solute concentration versus dimensionless time (left) and versus dimensionless distance (right).}
    \label{fig:comparison-solute}
    \end{figure}

We first numerically verify the integral representation \eqref{eq:integral-representation-final-solute} with the following initial and boundary conditions provided in \cite{de2019hybrid}:
\begin{align*}
    C_o(x) = 2 e^{-x}, \quad\quad C_{\text{Di}}(t) =B_0(t) + 1,
\end{align*}
where $B_0(t)$ is the Bessel function of the first kind of order 0.
Following the parameters estimated in \cite{genuchten2013exact}, we set $D_c = 11.4$ (m$^2$s$^{-1}$),$v_c = 0.426$ (ms$^{-1}$) and the average decay rate $\bar{\lambda}_c = 0.001$ (s$^{-1}$). We consider a time-varying decay rate $\lambda_c(t) = \bar{\lambda}_c(1+\sin(t)/2)$. 
Figure~\ref{fig:comparison-solute} shows an excellent agreement between the numerical evaluation of \eqref{eq:integral-representation-final-solute} and the fully numerical solution computed by MATLAB \texttt{pdepe}.

\subsubsection{Optimal control of solute transport}
\begin{figure}[tb]
    \begin{subfigure}{0.49\textwidth}
        \centering
        \includegraphics[width=0.95\textwidth]{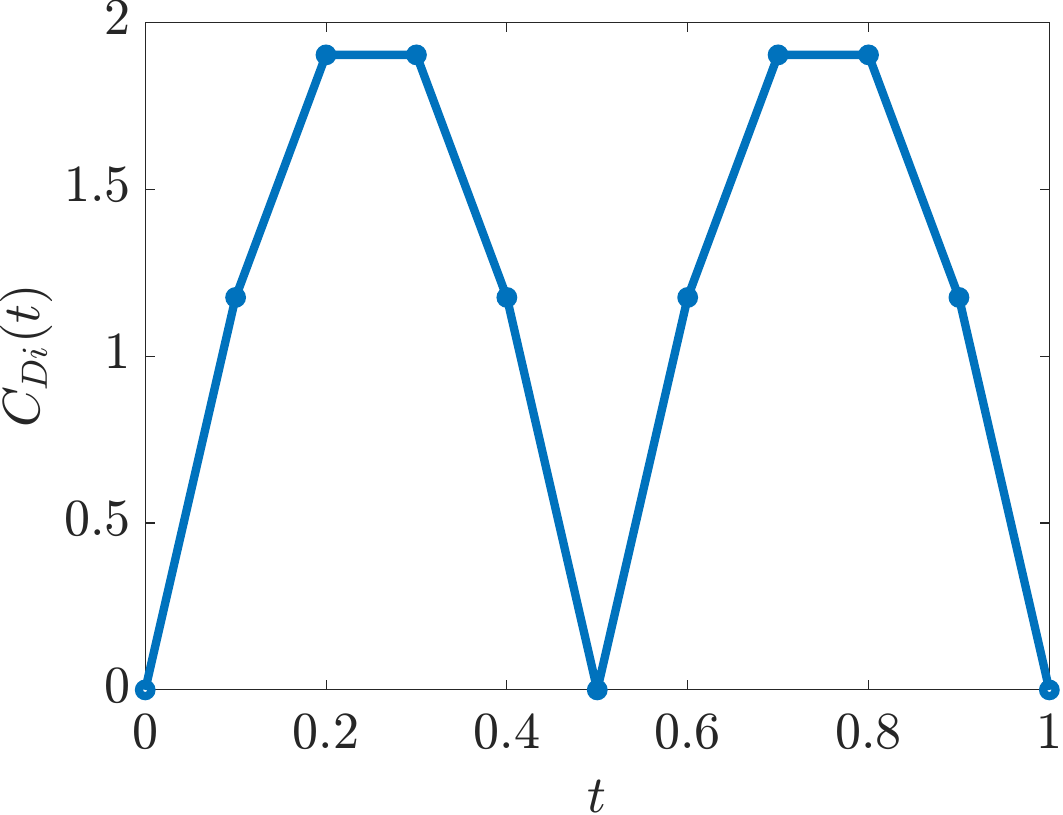}
        \caption{Boundary condition $C_{\text{Di}}(t)$}
        \label{fig:boundary-condition-solute-2}    
    \end{subfigure}\hfill
    \begin{subfigure}{0.49\textwidth}
        \centering
        \includegraphics[width=0.95\textwidth]{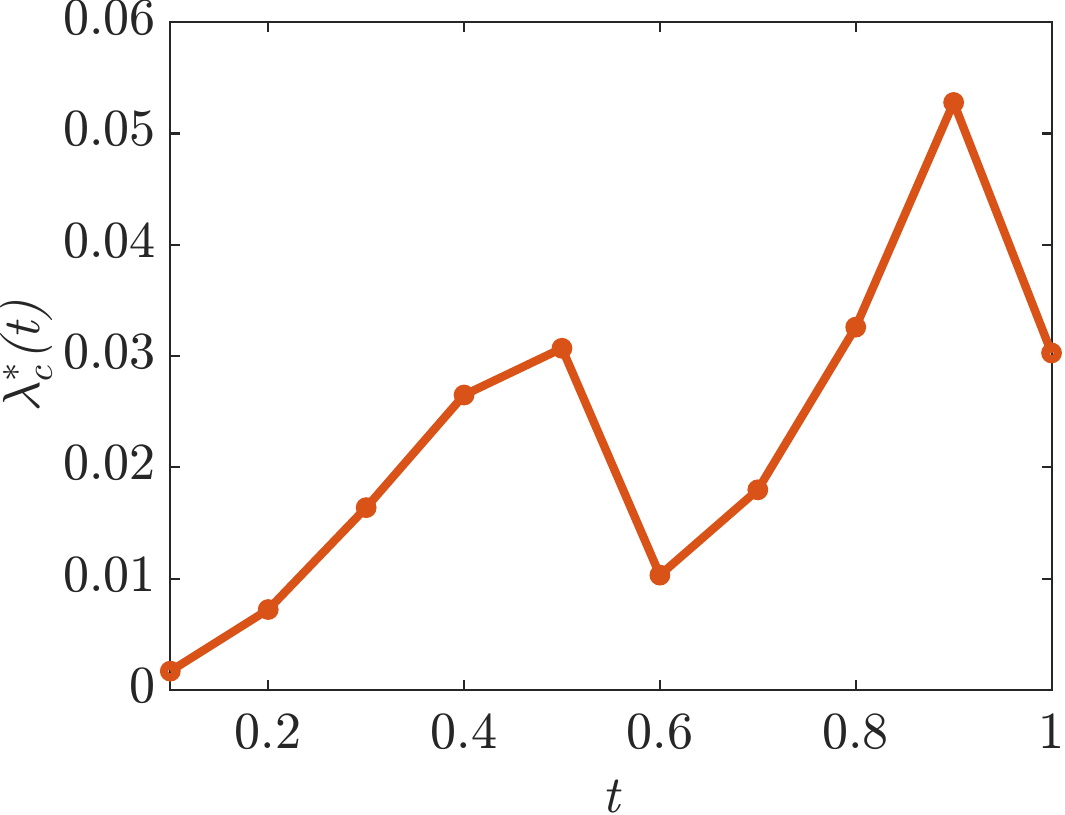}
        \caption{Optimal decay rate $\lambda^*_c(t)$}
        \label{fig:optimal-control-solute-2}
    \end{subfigure}
    \begin{subfigure}{0.49\textwidth}
        \centering
        \includegraphics[width=\textwidth]{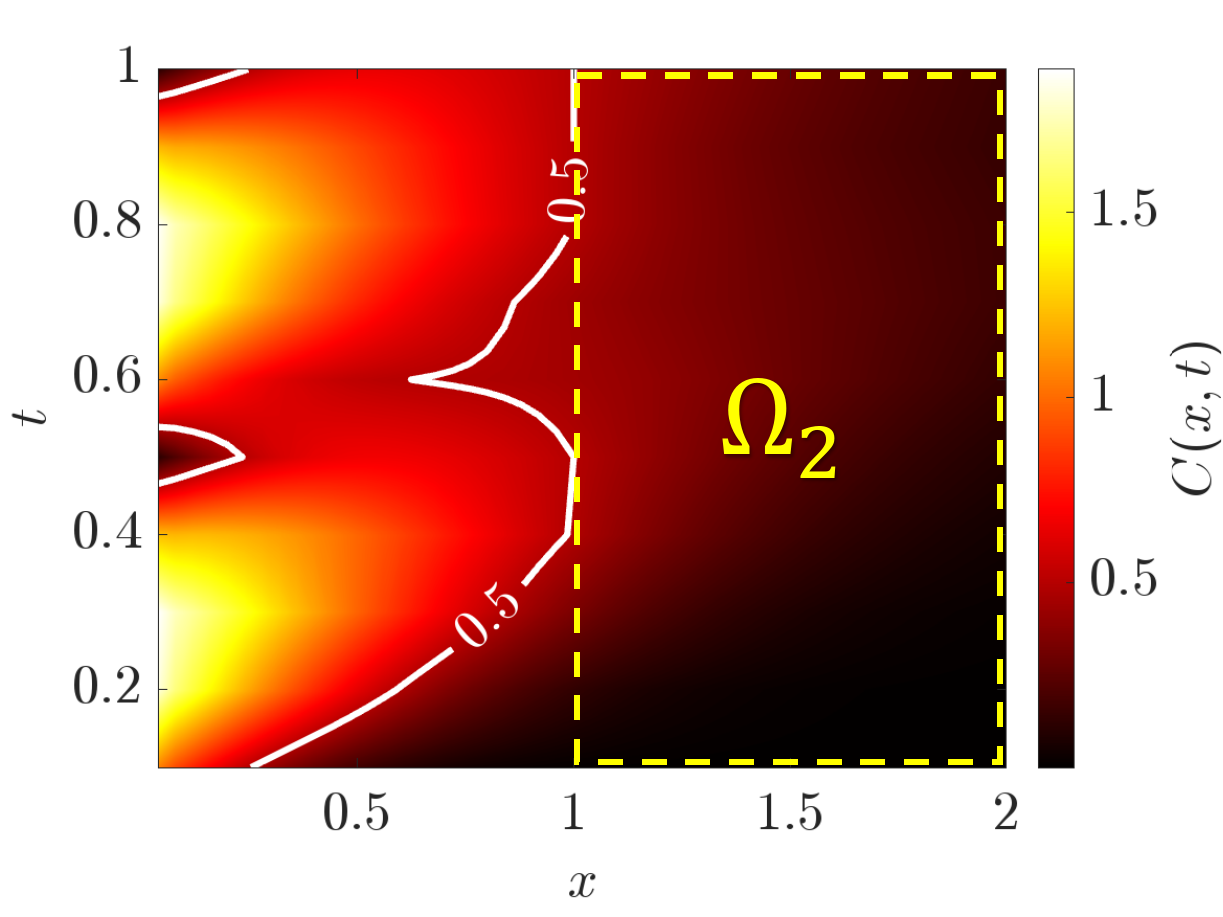}
        \caption{Solute concentration $C(x,t)$}
        \label{fig:state-contour-solute-2}
    \end{subfigure}
    \begin{subfigure}{0.49\textwidth}
        \centering
        \includegraphics[width=0.95\textwidth]{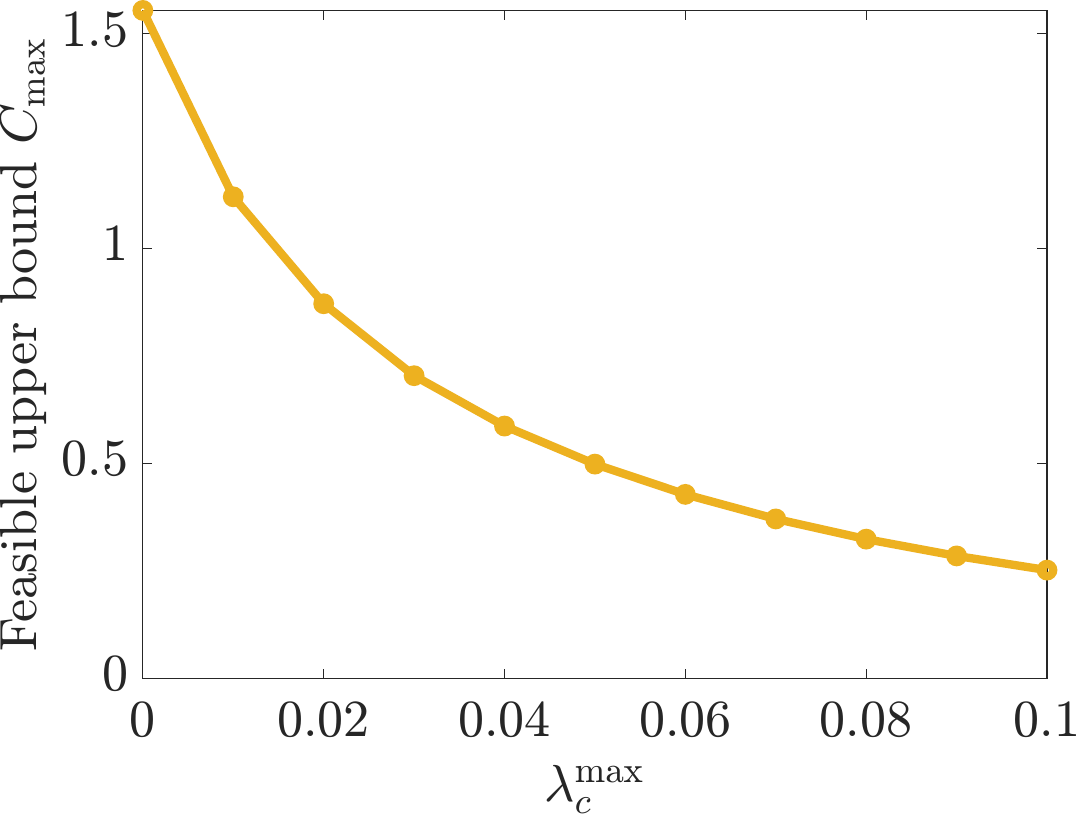}
        \caption{Feasible upper bound $C_{\max}$ }
        \label{fig:maximal-upper-bound-solute}
    \end{subfigure}
    \caption{Optimal control of solute transport with periodic boundary condition specified in (a). 
    The computed optimal control is shown in (b).
    The neutron flux $C(x,t)$ under optimal control is shown in (c) where $C(x,t)\leq 0.5$ is imposed in the region $\Omega_2 = \{x\geq 0.5, 0\leq t\leq 1\}.$
    (d) gives minimal $C_{\max}$ that is feasible for \eqref{eq:optimal-control-neutron} for different values of $\lambda_c^{\max}$.}
    \label{fig:solute-transport-control-2-periodic}
\end{figure}
\begin{figure}
    \begin{subfigure}{0.49\textwidth}
        \centering
        \includegraphics[width=0.95\textwidth]{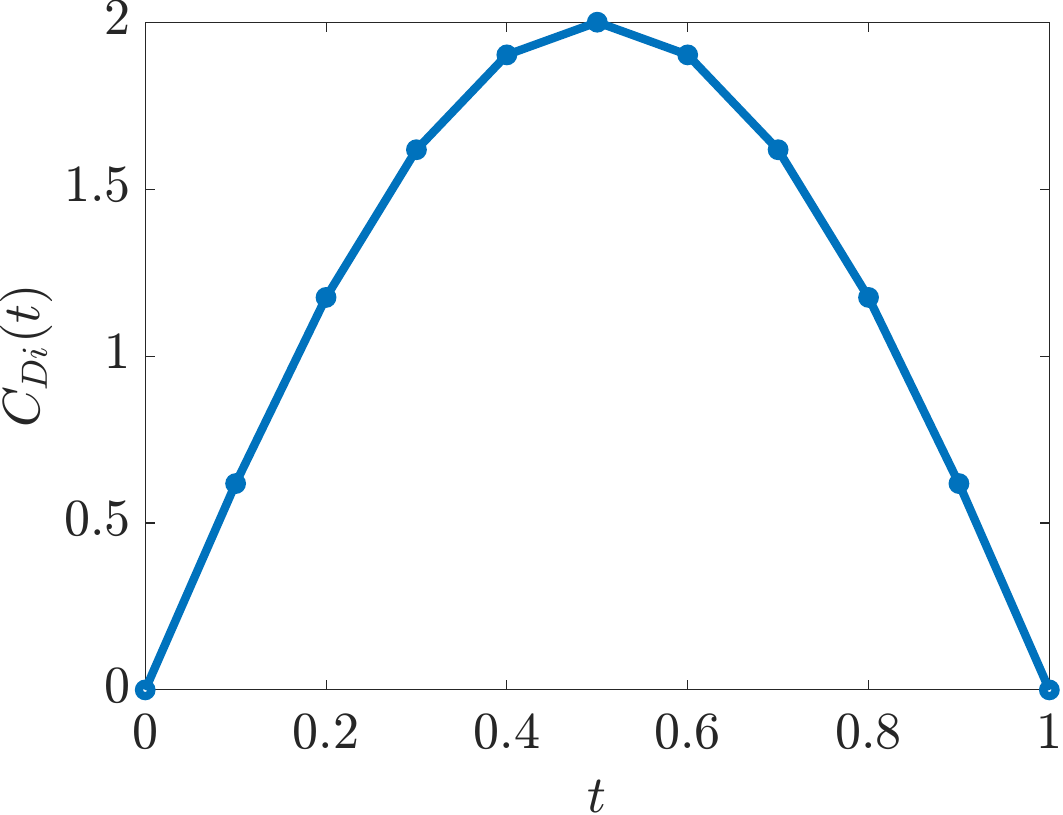}
        \caption{Boundary condition $C_{\text{Di}}(t)$}
        \label{fig:boundary-condition-period-1}    
    \end{subfigure}\hfill
    \begin{subfigure}{0.49\textwidth}
        \centering
        \includegraphics[width=0.95\textwidth]{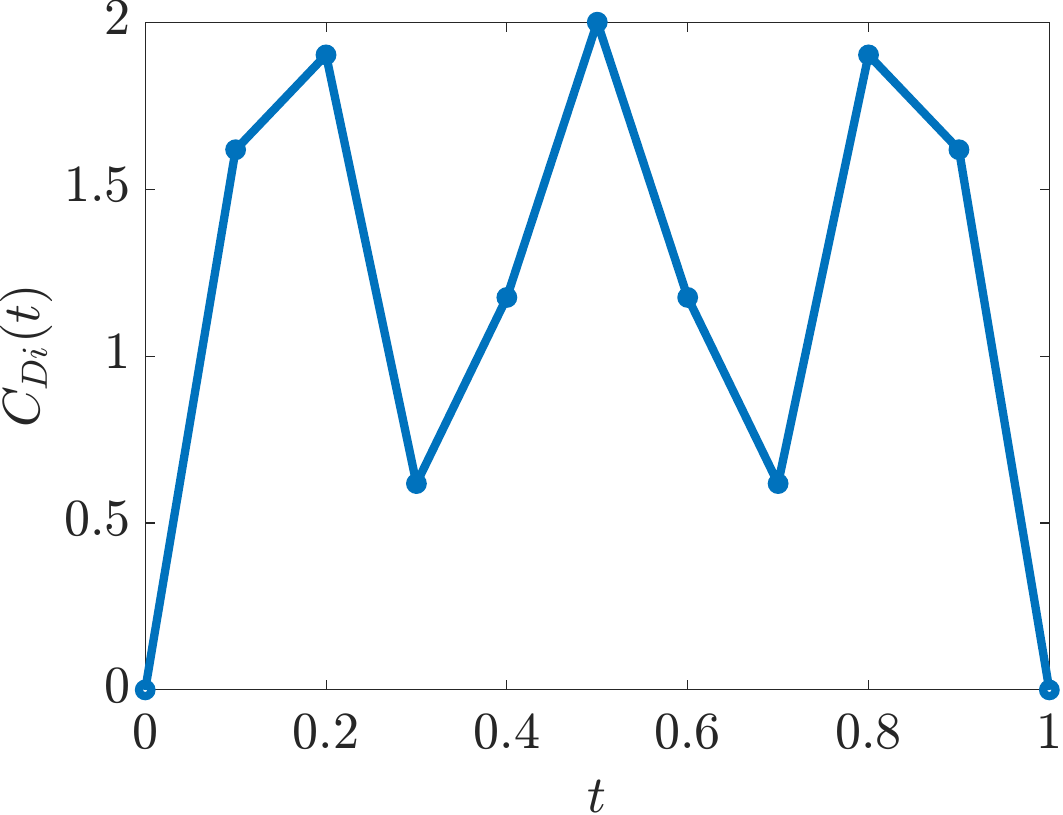}
        \caption{Boundary condition $C_{\text{Di}}(t)$}
        \label{fig:boundary-condition-period-4}    
    \end{subfigure}
    \begin{subfigure}{0.49\textwidth}
        \centering
        \includegraphics[width=0.95\textwidth]{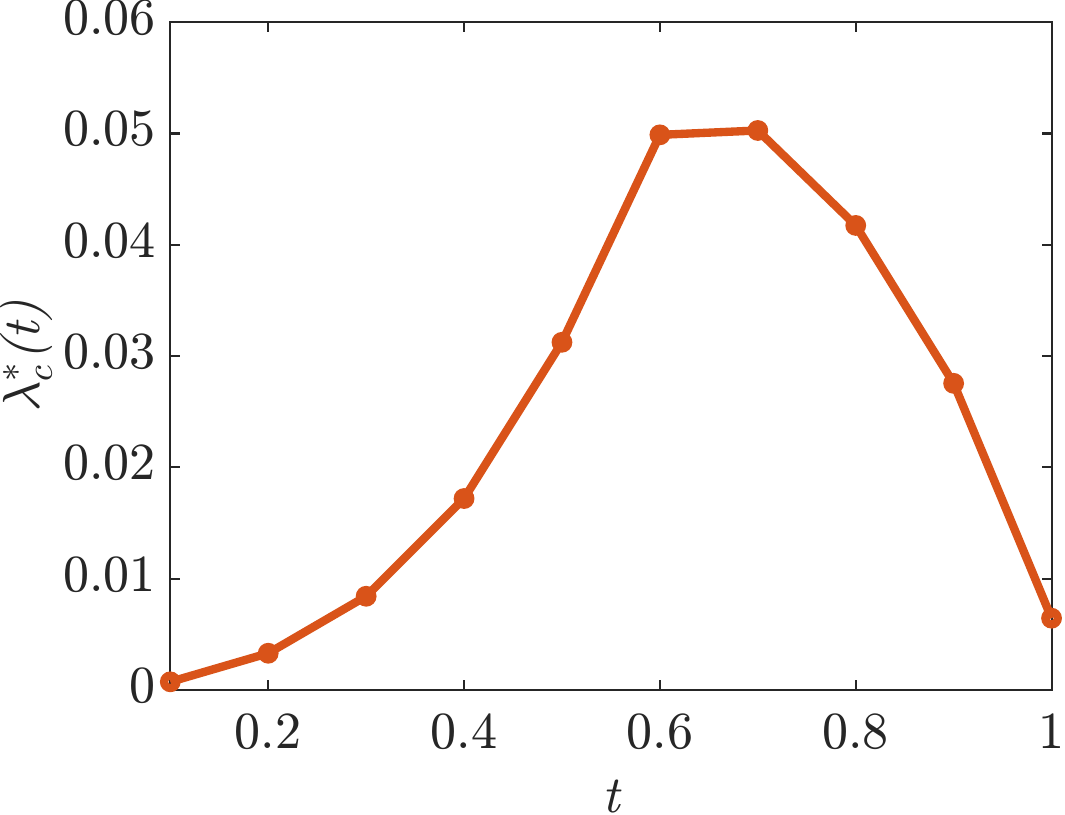}
        \caption{Optimal control $\lambda^*_c(t)$}
        \label{fig:optimal-control-solute-1}
    \end{subfigure}\hfill
    \begin{subfigure}{0.49\textwidth}
        \centering
        \includegraphics[width=0.95\textwidth]{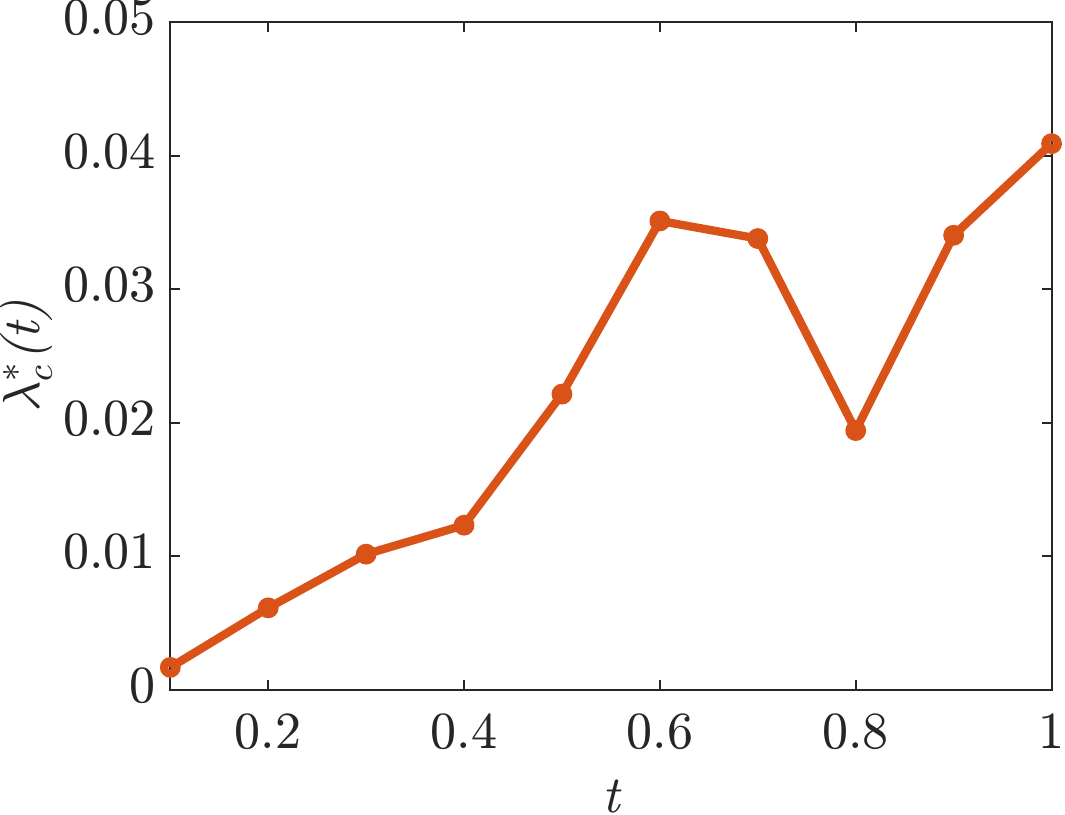}
        \caption{Optimal control $\lambda^*_c(t)$}
        \label{fig:optimal-control-solute-3}
    \end{subfigure}
    \begin{subfigure}{0.49\textwidth}
        \centering
        \includegraphics[width=\textwidth]{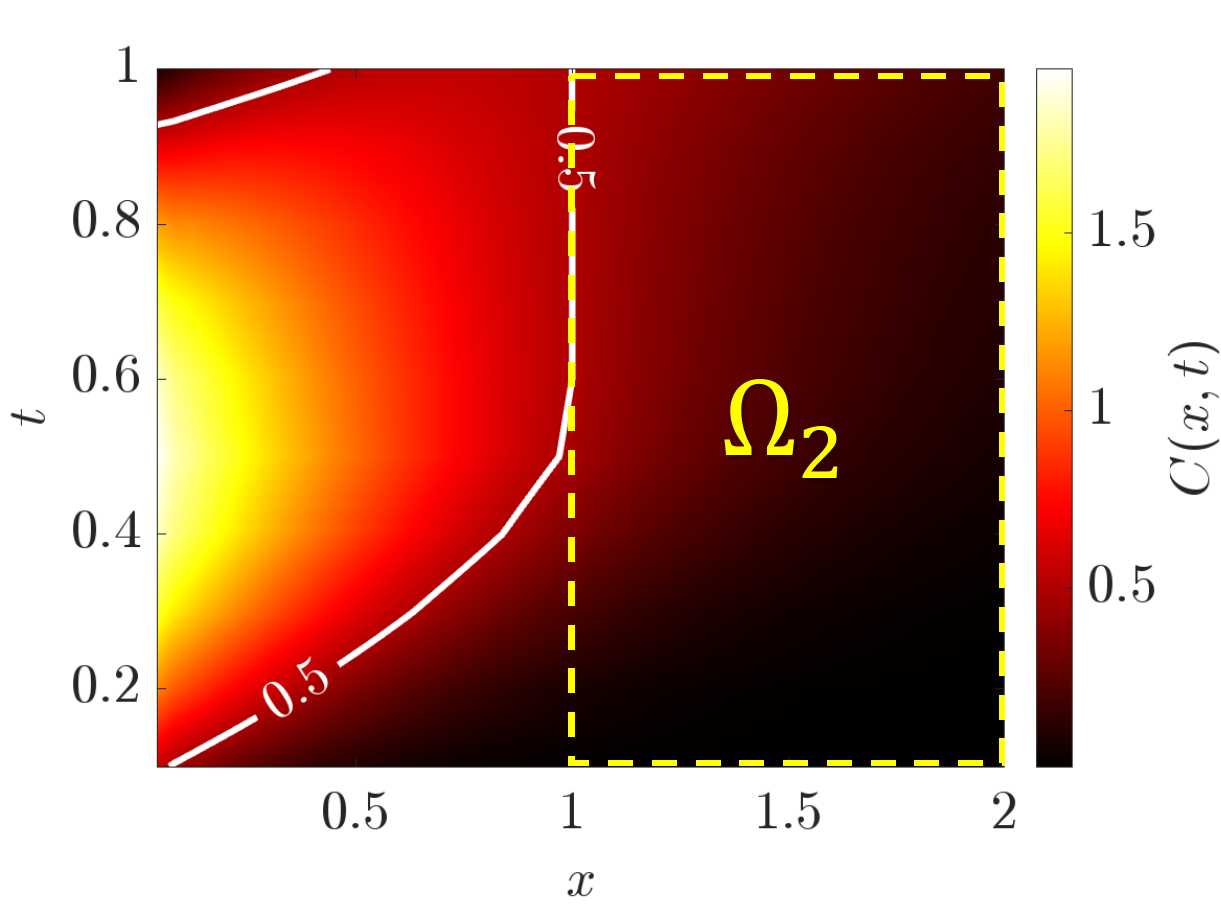}
        \caption{Solute concentration $C(x,t)$}
        \label{fig:state-contour-solute-period-1}
    \end{subfigure}\hfill
    \begin{subfigure}{0.49\textwidth}
        \centering
        \includegraphics[width=\textwidth]{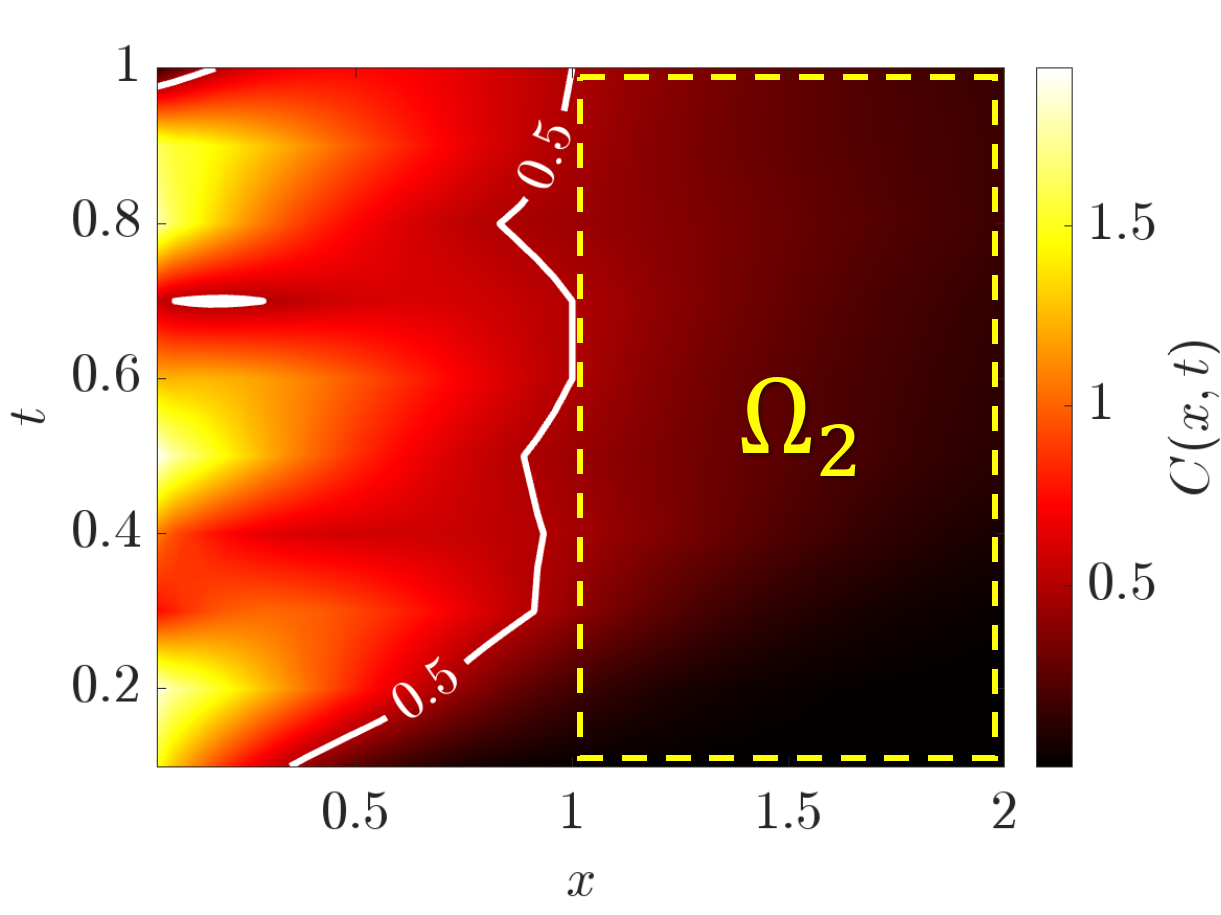}
        \caption{Solute concentration $C(x,t)$}
        \label{fig:state-contour-solute-period-3}
    \end{subfigure}
    \caption{Optimal control of solute transport with various boundary conditions specified in (a) and (b), respectively. The computed optimal control is shown in (c) and (d), respectively. The solute concentration under the computed control is shown in (e) and (f), respectively. The state constraint $C(x,t)\leq 0.5$ is imposed in the region $\Omega_2 = \{x\geq 0.5, 0\leq t\leq 1\}.$}
    \label{fig:solute-transport-control-periodic}
\end{figure}
In water treatment using catalytic converters \citep{heck2019catalytic}, one of the main goals is to reduce the concentration of contaminants below a threshold value after a certain position in the pipe while using a minimal amount of catalysts.
Therefore, we consider the following optimal control problem with an upper bound on the state:
\begin{equation}\label{eq:optimal-control-lower-bound}
    \begin{split}
        &\min_{0\leq\lambda_c\leq\lambda^{\max}_c} \quad \int_{0}^{T} \lambda_c^2(t) dt\\
        & \text{s.t.}\quad  C(x,t)\leq C_{\max},\quad (x,t) \in \Omega_2= \{x\geq L_1,0\leq t\leq T\},
    \end{split}
\end{equation}
where $T=1$, $C_{\max}=0.5$, $L_1=1$.
This represents the case when the solute (contaminant) concentration is required to be below $0.5$ after $x=1$.
The initial condition is set to $C_o(x)=0$, which means that there is initially no contaminant in the pipe.

Figure~\ref{fig:solute-transport-control-2-periodic} \zlcomment{illustrates the numerical result under} nonnegative periodic boundary condition $C_{\text{Di}}(t)=2|\sin(2\pi t)|$ \zlcomment{shown in Figure~\ref{fig:boundary-condition-solute-2}}.
This boundary condition corresponds to the case where the solute is injected into the conduit \zlcomment{from the left boundary} with a concentration that varies periodically \zlcomment{over} time.
Figure~\ref{fig:optimal-control-solute-2} shows that the computed optimal decay rate $\lambda_c^*(t)$ follows \zlcomment{a similar trend in the amplitude and period of} the boundary condition.
This indicates that optimal control adapts to the varying injection rate \zlcomment{at the left boundary} effectively.
\zlcomment{When the boundary value (injection rate) increases, the corresponding solute concentration value in the pipe also increases}, see the light yellow areas in Figure~\ref{fig:state-contour-solute-2}.
Hence, a higher decay rate is needed to keep the concentration below the upper bound $C_{\max}$, as shown in Figure~\ref{fig:optimal-control-solute-2}.

\zlcomment{To illustrate the effect of constraints}, Figure~\ref{fig:state-contour-solute-2} shows that the solute concentration value $C(x,t)$ is within the upper bound $C_{\max}=0.5$ for $(x,t)$ in the region $\Omega_2$. This is consistent with the state constraint in \eqref{eq:optimal-control-lower-bound}.
In other words, the catalytic converter is able to reduce the contaminant concentration below the threshold $C_{\max}$ for $x\geq1$ using the decay rate in Figure~\ref{fig:optimal-control-solute-2}.
\zlcomment{Similarly to Figure~\ref{fig:maximal-Upper-bound} for nuclear reactivity control,  
the orange curve in Figure~\ref{fig:maximal-upper-bound-solute} suggests the minimal values for $C_{\max}$ to ensure that optimal control problem \eqref{eq:optimal-control-lower-bound} is feasible.}
Physically, the orange curve in Figure~\ref{fig:maximal-upper-bound-solute} reflects different abilities to control the solute concentration value according to the ability of the catalytic converters to control the decay rate $\lambda_c$.

Figure~\ref{fig:solute-transport-control-periodic} illustrates the results under other nonnegative periodic boundary conditions shown in Figures~\ref{fig:boundary-condition-period-1}--\ref{fig:boundary-condition-period-4}.
It can be seen from Figures~\ref{fig:optimal-control-solute-1} --\ref{fig:optimal-control-solute-3} that the computed control adapts to the patterns of the boundary conditions, such as the number of periods.
This is consistent with Figure~\ref{fig:solute-transport-control-2-periodic}.
The light yellow areas in Figures~\ref{fig:state-contour-solute-period-1}--\ref{fig:state-contour-solute-period-3} indicate high concentration values, which are related to the peaks of the boundary conditions in Figures~\ref{fig:boundary-condition-period-1} and \ref{fig:boundary-condition-period-4}.
\zlcomment{The concentration $C(x,t)$ in $\Omega_2$ is within the upper bound} $C_{\max}=0.5$ in both Figures~\ref{fig:state-contour-solute-period-1} and \ref{fig:state-contour-solute-period-3}, similar to the previous case in Figure~\ref{fig:state-contour-solute-2}.

\section{Conclusions}
\label{sec:conclusion}
We proposed an \zlcomment{optimize-then-discretize} computational framework for solving constrained bilinear optimal control problems for second-order linear evolution PDEs with both state and control constraints.
Existing approaches \zlcomment{lack global convergence guarantee} and have not considered state constraints due to the complexity of control-to-state mappings arising from bilinearity.
Our framework derives an integral representation of the PDE solution using the unified transform method, which can be seen as an explicit expression for the control-to-state mapping used in the literature.
Such an integral representation gives rise to explicit expressions for derivatives with respect to the control variable that are easy to evaluate numerically.
\zlcomment{The integral representation is used to replace the PDE constraint, which results in an equivalent reformulation of the optimal control problem and circumvents the difficulties associated with PDE analysis.} 
Then, the KKT conditions for the reformulated problem are derived and discretized into \zlcomment{a system of finite-dimensional smooth nonlinear} equations that can be solved by existing algorithms \zlcomment{with Q-quadratic convergence}.
Our framework preserves the PDE relation in continuous space and time, unlike conventional methods that discretize the PDE to solve for optimal control. 
We applied the framework to two application problems: nuclear reactivity control and water quality treatment in a reactor.
The computational results illustrate the effectiveness of the framework for these problems.

Future works include extending the framework to more complex systems, such as higher dimensions or network systems. 
In addition to second-order linear evolution PDEs, the framework can also be applied to other types of PDEs where the unified transform method is applicable, such as axial load control for vibrating beams described by the linear wave equation \citep{ball1982controllability,khapalov2010controllability}.
\zlcomment{Furthermore, we have only considered first-order optimality conditions in this work. Using our framework to investigate second-order optimality conditions will be pursued in the future. }

\section*{CRediT authorship contribution statement}
\textbf{Zhexian Li}: Conceptualization, Methodology, Software, Validation, Formal analysis, Investigation, Data curation, Writing - Original Draft, Visualization. 
\textbf{Felipe P. J. de Barros}: Conceptualization, Methodology, Validation, Formal analysis, Investigation, Writing - review $\&$ editing, Visualization, Supervision, Funding acquisition. 
\textbf{Ketan Savla}: Conceptualization, Methodology, Validation, Formal analysis, Investigation, Writing - review $\&$ editing, Visualization, Supervision, Funding acquisition. 

\section*{Data availability}

The code and data supporting the findings of this study are openly available on GitHub at \url{https://github.com/zhexianli-usc/Bilinear-control-of-advection-diffusion-equations}.

\section*{Acknowledgments}
\zlcomment{The authors thank the Editor-in-Chief, Dr. Reali, and the anonymous Associate Editor for giving us the opportunity to revise the paper.
The authors also thank the two anonymous reviewers for their constructive feedback that helped significantly improve the manuscript. 
}The authors acknowledge the financial support provided by the National Science Foundation Grant number 1654009.

\appendix
\section{Integral representations of the PDE solutions}
\label{sec:pde-representation}
In this appendix, we present a general procedure for obtaining integral representations of the solutions to \eqref{eq:general-second-order}.
Then we show how to obtain explicit expressions \eqref{eq:integral-representation-final-neutron} and \eqref{eq:integral-representation-final-solute} for the neutron flux and the solute concentration as examples, respectively.
Previously, the unified transform method has mainly been applied to PDEs with constant coefficients and has not considered time-varying coefficients \citep{deconinck2014method,fokas2023modern}.
Here we extend the method to \eqref{eq:general-second-order} with time-varying coefficient $u(t)$.

\subsection{Integral representation}
We first introduce the following Fourier transform pair:
\begin{align}
    \label{eq:fourier-transform}
    \hat{\psi}(k,t) &= \int_{0}^{\infty} \psi(x,t) e^{-ikx} dx,\quad\imag[k]\leq0,\\
   \psi(x,t) &= \int_{-\infty}^{\infty} \hat{\psi}(k,t) e^{ikx} \frac{dk}{2\pi}, \quad 0<x<\infty,
\end{align}
where $i$ denotes the imaginary unit and $k\in\mathbb{C}$.

We will obtain solutions of \eqref{eq:general-second-order} subject to two distinct boundary conditions.
Let $\psi_{\text{Di}}(t)$ and $\psi_{\text{Ne}}(t)$ denote the Dirichlet and Neumann boundary values, respectively, i.e.,
\begin{align*}
     \psi_{\text{Di}}(t) &= \psi(0,t),\quad t\geq 0,\\     
    \psi_{\text{Ne}}(t) &= \frac{\partial \psi}{\partial x}(0,t),\quad t\geq 0.
\end{align*}
Substituting the Fourier transform pair into \eqref{eq:general-second-order} and using integration by parts, we obtain the following ODE,
\begin{multline}\label{eq:transformed-ode}
    \frac{\partial \hat{\psi}(k,t)}{\partial t} \\= -[\alpha k^2+ivk+u(t)]\hat{\psi}(k,t) - \alpha \psi_{\text{Ne}}(t)-(i\alpha k-v) \psi_{\text{Di}}(t) + \hat{f}(k,t),
\end{multline}
with the initial condition $\hat{\psi}(k,0) = \hat{\psi}_o(k)$, where $\hat{\psi}_o(k)$ and $\hat{f}(k,t)$ are the Fourier transform \eqref{eq:fourier-transform} applied to the initial condition $\psi_o(x)$ and the forcing term $f(x,t)$, respectively.
Solving the ODE, see \eqref{eq:transformed-ode}, gives
\begin{equation}\label{eq:global-relation}
    e^{\omega(k,t,u)}\hat{\psi}(k,t) = \hat{\psi}_o(k) -\alpha\tilde{\psi}_{\text{Ne}}(k,t,u) -(i\alpha k-v)\tilde{\psi}_{\text{Di}}(k,t,u) + \tilde{f}(k,t,u),
\end{equation}
where 
\begin{equation}\label{eq:global-relation-terms}
\begin{split}
    &\omega(k,t,u) = (\alpha k^2 + ivk)t +\tilde{u}(t), \quad \tilde{u}(t) = \int_{0}^{t} u(\tau)d\tau,\\
    &\tilde{\psi}_{\text{Ne}}(k,t,u) = \int_{0}^{t} e^{\omega(k,\tau,u)}\psi_{\text{Ne}}(\tau)d\tau,\, \tilde{\psi}_{\text{Di}}(k,t,u) = \int_{0}^{t} e^{\omega(k,\tau,u)}\psi_{\text{Di}}(\tau)d\tau,  \\
    &\tilde{f}(k,t,u) = \int_{0}^{t} e^{\omega(k,\tau,u)}\hat{f}(k,\tau)d\tau.
\end{split}
\end{equation}
Employing the inverse Fourier transform in \eqref{eq:global-relation}, we find
\begin{multline}\label{eq:integral-representation-real-line}
    \psi(x,t,u) = \int_{-\infty}^\infty e^{ikx-\omega(k,t,u)} \left[\hat{\psi}_o(k) + \tilde{f}(k,t,u)\frac{dk}{2\pi}\right] \\ + \int_{-\infty}^\infty e^{ikx-\omega(k,t,u)} \left[ - \alpha\tilde{\psi}_{\text{Ne}}(k,t,u) -(i\alpha k-v)\tilde{\psi}_{\text{Di}}(k,t,u) \right] \frac{dk}{2\pi}.
\end{multline}
In \eqref{eq:integral-representation-real-line}, we obtain an integral representation for $\psi(x,t)$ that involves both the Neumann boundary value $\psi_{\text{Ne}}(t)$ and the Dirichlet boundary value $\psi_{\text{Di}}(t)$.
However, only one of the boundary values is given, e.g., the Neumann boundary value is given for \eqref{eq:diffusion-equation-neutron} and the Dirichlet boundary value is given for \eqref{eq:solute-transport-equation}. 
Next, we show how to eliminate the unknown boundary value in \eqref{eq:integral-representation-real-line}.

\subsection{Contour deformation}
\begin{figure}
    \centering
    \includegraphics[width = 0.5\textwidth]{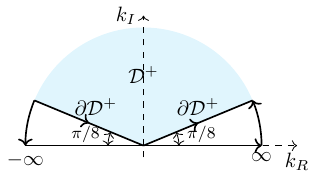}
    \caption{Contour deformation from the real line to the contour $\partial\mathcal{D}^+$ in the upper half of the complex plane.
    $k_R$ and $k_I$ are the real and imaginary parts of $k$, respectively.}
    \label{fig:contour-deformation}
\end{figure}
First, we deform the integrals in \eqref{eq:integral-representation-real-line} from the real line to a contour in the upper half of the complex plane. 
The deformed contour is chosen as $\partial \mathcal{D}^+:=\{k\in\mathbb{C}^+:k=|k|e^{i\theta},\theta = \pi/8 \text{ or }7\pi/8\}$ as shown in Figure \ref{fig:contour-deformation}. 
Note that the argument $\theta$ can take arbitrary value in $(0,\pi/4)$ or $(3\pi/4,\pi)$, and the choice of $\theta$ does not affect the final result.
Following the unified transform method, the requirement for contour deformation is that the integrand in \eqref{eq:integral-representation-real-line} is analytic and decays sufficiently fast as $k\to\infty$ in $\mathbb{C}^+\setminus\mathcal{D}^+$, i.e., the white region between the real line and $\partial\mathcal{D}^+$ as shown in Figure \ref{fig:contour-deformation}.
The contour deformation follows from similar arguments in \cite{deconinck2014method,de2019hybrid,fokas2023modern}, and we will only show that the integrand satisfies the analytic and decaying requirements.

Suppose that the initial and boundary conditions and the forcing term $f$ are all analytic. 
Then, the integrand in \eqref{eq:integral-representation-real-line} is analytic since the integrand is a composition of analytic functions.
Next, we show that the integrand decays exponentially as $k\to\infty$ in $\mathbb{C}^+\setminus\mathcal{D}^+$.
We rewrite the integrand of the second integral in \eqref{eq:integral-representation-real-line} as $e^{ikx}G(k;t)$ and show that $G(k;t)$ decays exponentially as $k\to\infty$ in $\mathbb{C}^+\setminus\mathcal{D}^+$.
Recall from \eqref{eq:global-relation-terms} that $\tilde{\psi}_{\text{Ne}}, \tilde{\psi}_{\text{Di}}$ all contain the exponential term $e^{\omega(k,\tau,u)}$ where $0<\tau<t$. 
Therefore, the exponential term in $G(k;t)$ is $\exp[-(\omega(k,t,u)-\omega(k,\tau,u))]$.
As $k\to\infty$ in $\mathbb{C}^+\setminus\mathcal{D}^+$, the leading term in the exponent is $-\alpha(t-\tau)k^2$. 
Since $\text{Re}[k^2]>0$ in $\mathbb{C}^+\setminus\mathcal{D}^+$ and $t-\tau>0$, $\exp(-\alpha(t-\tau)k^2)$ decays exponentially, and thus $G(k;t)$ decays exponentially in $\mathbb{C}^+\setminus\mathcal{D}^+$.
Therefore, we obtain the following integral representation after contour deformation:
\begin{multline}\label{eq:integral-representation-D-+}
    \psi(x,t,u) = \int_{-\infty}^\infty e^{ikx-\omega(k,t,u)} \left[\hat{\psi}_o(k)+ \tilde{f}(k,t,u)\right]\frac{dk}{2\pi} \\
    + \int_{\partial\mathcal{D}^+} e^{ikx-\omega(k,t,u)}\left[ - \alpha\tilde{\psi}_{\text{Ne}}(k,t,u) -(i\alpha k-v)\tilde{\psi}_{\text{Di}}(k,t,u)\right]\frac{dk}{2\pi}.
\end{multline}
\subsection{Elimination of the unknowns}
For the two problems considered in Sections~\ref{sec:nuclear-reactor}--\ref{sec:ADE}, we show how to eliminate the unknowns in \eqref{eq:integral-representation-D-+}, respectively.
Since $\omega(k,t,u)$ in \eqref{eq:global-relation-terms} is a second-order polynomial in $k$, there exists a nontrivial $\nu(k)$ such that $\nu(k)\neq k$ and $\omega(k,t,u) = \omega(\nu(k),t,u)$.
Solving the equation $\omega(k,t,u)=\omega(\nu(k),t,u)$, we find $\nu(k) = -k-iv/\alpha$.
Recall from \eqref{eq:global-relation-terms} that the dependence of $\tilde{\psi}_{\text{Ne}},\tilde{\psi}_{\text{Di}}$ on $k$ is through the function $\omega(k,t,u)$, and thus these two functions remain invariant under the transformation $k\to\nu(k)$.
Substituting $\nu(k)$ into \eqref{eq:global-relation}, we obtain
\begin{equation}\label{eq:global-relation-transformed}
    \begin{aligned}
        & e^{\omega(k,t,u)}\hat{\psi}(\nu(k),t,u) \\
        =\ &  \hat{\psi}_o(\nu(k)) -\alpha\tilde{\psi}_{\text{Ne}}(k,t,u) -(i\alpha\nu(k)-v)\tilde{\psi}_{\text{Di}}(k,t,u) + \tilde{f}(\nu(k),t,u) \\
        =\ & \hat{\psi}_o(\nu(k)) -\alpha\tilde{\psi}_{\text{Ne}}(k,t,u) +i\alpha k\tilde{\psi}_{\text{Di}}(k,t,u) + \tilde{f}(\nu(k),t,u).
    \end{aligned}
\end{equation}
From \eqref{eq:global-relation-transformed}, we have
\begin{multline}\label{eq:g1-elimination}
    \alpha\tilde{\psi}_{\text{Ne}}(k,t,u) \\= \hat{\psi}_0(\nu(k))+ i\alpha k\tilde{\psi}_{\text{Di}}(k,t,u) -  e^{\omega(k,t,u)}\hat{\psi}(\nu(k),t,u) + \tilde{f}(\nu(k),t,u),
\end{multline}
and
\begin{multline}\label{eq:g0-elimination}
   i\alpha k \tilde{\psi}_{\text{Di}}(k,t,u)\\ = -\hat{\psi}_0(\nu(k))+\alpha \tilde{\psi}_{\text{Ne}}(k,t,u) +  e^{\omega(k,t,u)}\hat{\psi}(\nu(k),t,u) - \tilde{f}(\nu(k),t,u).
\end{multline}

For \eqref{eq:dimensionless-solute-transport-equation} with the Dirichlet boundary condition, substituting \eqref{eq:g1-elimination} into \eqref{eq:integral-representation-D-+}, we find 
\begin{multline}\label{eq:integral-representation-final}
    \psi(x,t,u)=\int_{-\infty}^\infty e^{ikx-\omega(k,t,u)} \left[\hat{\psi}_o(k)+ \tilde{f}(k,t,u)\right]\frac{dk}{2\pi} \\ 
    - \int_{\partial\mathcal{D}^+}\!\!\! e^{ikx-\omega(k,t,u)}  \left[ \hat{\psi}_o(\nu(k))\! +\! (2i\alpha k-v)\tilde{\psi}_{\text{Di}}(k,t) + \tilde{f}(\nu(k),t,u)\right]\frac{dk}{2\pi},
\end{multline}
plus the following integral that vanishes:
\begin{equation*}
    \int_{\partial\mathcal{D}^+} e^{ikx}\hat{\psi}(\nu(k),t,u)\frac{dk}{2\pi}.
\end{equation*}
The above integral vanishes following similar arguments provided in \cite{deconinck2014method,fokas2023modern} due to the fact that $\hat{\psi}(\nu(k),t,u)$ decays exponentially in $\mathbb{C}^+$.
This is because $\exp[-i\nu(k)x] = \exp[(ik - v/\alpha)x]$ decays exponentially since $\exp[ikx] = \exp[ik_R x - k_I x]$ using the definition $k= k_R + ik_I$.
Substituting the coefficients of \eqref{eq:dimensionless-solute-transport-equation}, we find that \eqref{eq:integral-representation-final} reduces to \eqref{eq:integral-representation-final-solute}.

Similarly, for \eqref{eq:diffusion-equation-neutron} with the Neumann boundary condition, substituting \eqref{eq:g0-elimination} into \eqref{eq:integral-representation-D-+}, we obtain
\begin{multline}\label{eq:integral-representation-final-neumann}
    \psi(x,t,u) = \int_{-\infty}^\infty e^{ikx-\omega(k,t,u)} \left[\hat{\psi}_o(k)+ \tilde{f}(k,t,u)\right]\frac{dk}{2\pi} \\
    +\int_{\partial\mathcal{D}^+} e^{ikx-\omega(k,t,u)}
    \left[\hat{\psi}_o(-k) - 2\alpha\tilde{\psi}_{\text{Ne}}(k,t,u) + \tilde{f}(-k,t,u)\right]\frac{dk}{2\pi}.
\end{multline}
Substituting the coefficients of \eqref{eq:diffusion-equation-neutron}, we find that \eqref{eq:integral-representation-final-neumann} reduces to \eqref{eq:integral-representation-final-neutron}.

\section{Derivation of optimality conditions}
\label{sec:appendix-directional-derivative}
In this appendix, we show how to derive the necessary conditions \eqref{eq:kkt-infinite-dim}--\eqref{eq:derivative-psi}.
We first introduce the concepts of directional derivative and differentiability.
Then, we derive the optimality conditions using the KKT conditions for infinite-dimensional optimization with differentiable objectives and constraints \cite[Section 2.5.5]{hinze2008optimization}.

\subsection{Directional derivative}
\label{sec:directional-derivative}
The advantage of using integral representations, e.g., \eqref{eq:integral-representation-final-neutron} and \eqref{eq:integral-representation-final-solute}, is that we can derive explicit expressions for the directional derivatives of $\psi(x,t,u)$ with respect to the control variable $u$. 
We first give a definition of directional derivative and then provide explicit expressions for the directional derivatives of \eqref{eq:integral-representation-final-neutron} and \eqref{eq:integral-representation-final-solute}, respectively.
Following \cite[Definition 1.29]{hinze2008optimization}, directional derivative of a functional $F$ at a point $y$ in the direction $h$ is defined as
\begin{equation}
    \label{eq:gates-derivative}
    \begin{split}
        dF(y;h) &= \lim_{\epsilon\to 0^+}\frac{F(y+\epsilon h)-F(y)}{\epsilon} 
        = \left.\frac{d}{d\epsilon} F(y+\epsilon h)\right|_{\epsilon=0}.\\
    \end{split}
\end{equation}
If the directional derivative exists for all $h$ and the operator $\mathfrak{D}F(y):h\to dF(y;h)$ is bounded and linear, then $F$ is Gâteaux differentiable at $y$. 
Moreover, $F$ is Fréchet differentiable at $y$ if the following condition holds:
\begin{equation*}
    \label{eq:directional-derivative-condition}
    \lim_{\|h\|\to0^+}\frac{\|F(y+h) - F(y) - dF(y;h)\|}{\|h\|} = 0,
\end{equation*}
where $\|\cdot\|$ denotes a norm for the space of square-integrable functions.
A functional $F$ is said to be Fréchet differentiable if it is Fréchet differentiable everywhere.
It can be seen that $\phi$ in \eqref{eq:integral-representation-final-neutron} and $C$ in \eqref{eq:integral-representation-final-solute} depend on their corresponding control variables $\Sigma_a$ and $\lambda_c$ through the exponential terms $\exp[\int_{0}^{t}\Sigma_a(\tau) d\tau]$ and $\exp[\int_{0}^{t}\lambda_c(\tau) d\tau]$, respectively. Therefore, $\phi$ and $C$ are Fréchet differentiable since \eqref{eq:integral-representation-final-neutron} and \eqref{eq:integral-representation-final-solute} are compositions of Fréchet differentiable functionals and functions. 
After evaluating \eqref{eq:gates-derivative},
directional derivatives of \eqref{eq:integral-representation-final-neutron} and \eqref{eq:integral-representation-final-solute} are given by the following formulas. For simplicity, we omit the dependence on $x$ and $t$:
\begin{multline}\label{eq:directional-derivative-neutron}
        d\phi(\Sigma_a;h) =-\int_{0}^{t} d\tau\ h(\tau)  \int_{-\infty}^{\infty} e^{ikx - \omega_\phi(k,t,\Sigma_a)} D_n \hat{\phi}_o(k) \frac{dk}{2\pi} \\ -\int_{0}^{t} d\tau\ h(\tau)  \int_{\partial\mathcal{D}^+} e^{ikx - \omega_\phi(k,t,\Sigma_a)} D_n\left[ \hat{\phi}_o(-k)  - \frac{2}{D_n} \hat{\xi}_{\text{Ne}}(k,\tau,\Sigma_a) \right] \frac{dk}{2\pi},
\end{multline}
\begin{multline}\label{eq:directional-derivative-solute}
        dC(\lambda_c;h) = -\int_{0}^{t} d\tau\ h(\tau)\int_{-\infty}^{\infty}e^{ikx - \omega_c(k,t,\lambda_c)}\frac{D_c}{v_c^2}\hat{C}_o(k) \frac{dk}{2\pi}+\\
        \int_{0}^{t}\!\! d\tau\, h(\tau) \!\!\int_{\partial\mathcal{D}^+}\!\!\! e^{ikx - \omega_c(k,t,\lambda_c)}\frac{D_c}{v_c^2}\left[ \hat{C}_o(-k-i)\! +\! (2ik-1)\hat{C}_{\text{Di}}(k,\tau,\lambda_c)\right]\! \frac{dk}{2\pi} .
\end{multline}
Directional derivatives \eqref{eq:directional-derivative-neutron} and \eqref{eq:directional-derivative-solute} can be compactly written in the form 
\begin{equation}\label{eq:directional-derivative-general}
    d\psi(u;h) = \int_{0}^{t} d\tau\ h(\tau) \frac{\delta\psi}{\delta u}(x,t,u,\tau),
\end{equation}
where the explicit expression for $\frac{\delta\psi}{\delta u}$ for the two states $\phi$ and $C$ can be found in \eqref{eq:derivative-psi}.
% Equivalently, $$\frac{\delta\psi}{\delta u}(x,t,u,\tilde{\tau}) = d\psi(u;\delta(\tau-\tilde{\tau})),0\leq\tilde{\tau}\leq T$$
% Note that this is consistent with the expression \eqref{eq:functional-derivative-lagrangian}.
Similarly, directional derivative of the objective functional $J(u)$ can be written as
\begin{equation}
    \label{eq:directional-derivative-objective}
            dJ(u;h) = \int_0^T d\tau\ h(\tau) \frac{\delta J}{\delta u}(u,\tau),
\end{equation}
where the explicit expression for $\frac{\delta J}{\delta u}$ can be found in \eqref{eq:kkt-infinite-lagrangian}.

\subsection{Infinite-dimensional KKT conditions}
\label{sec:kkt}
Following \cite[Section 2.5.5]{hinze2008optimization}, we use the KKT conditions to derive necessary conditions of optimality for the reformulated problem \eqref{eq:optimal-control-reformulate}.
Using \eqref{eq:kkt-infinite-constraints}, the constraints in \eqref{eq:optimal-control-reformulate} can be rewritten as $g_j(x,t,u)\leq 0$ on $\Omega_j, j=1,\ldots,4$.
Let $\lambda_j$ denote a \zlcomment{square-integrable} function defined on $\Omega_j,j=1,\ldots,4$, also known as the Lagrange multiplier associated with $g_j$. 
Suppose that the interior of the constraints is nonempty, i.e., there exists $\bar{u}$ such that $g_j(x,t,\bar{u})<0$ on $\Omega_j$ for all $j=1,\ldots,4$. Then, for every optimal $u^*$ of \eqref{eq:optimal-control-reformulate}, there exist Lagrange multipliers $\lambda^*_j,j=1\ldots,4$ such that $(u^*,\lambda^*_1,\ldots,\lambda^*_4)$ satisfies the following KKT conditions
\begin{equation}\label{eq:kkt-conditions}
    \begin{split}
        &dJ(u^*;h) + \sum_{j=1}^{4}dG_j(u^*;h) = 0,\ \text{for all \zlcomment{square-integrable} $h$ on } [0,T], \\
        &g_j(x,t,u^*)\leq 0\text{ and } \lambda_j^*(x,t)\geq 0\text{ on }\Omega_j, \quad j=1,\ldots,4, \\
        &G_j(u^*,\lambda_j^*) = 0, \quad j=1,\ldots,4, \\
    \end{split}
\end{equation}
where
\begin{equation*}
\begin{split}
    G_j(u,\lambda_j) &= \int_{0}^{T}\int_{L_1}^{L_2} \lambda_j(x,t)g_j(x,t,u)dx \ dt,\quad j = 1,2, \\
    G_j(u,\lambda_j) &= \int_{0}^{T} \lambda_j(t)g_j(x,t,u)dt,\quad  j=3,4. \\
\end{split}
\end{equation*}
Note that we omit the dependence of $dG_j$ on $\lambda_j$ in \eqref{eq:kkt-conditions} to emphasize that the directional derivative is taken with respect to $u$.

Following the expressions of $g_j$ in \eqref{eq:kkt-infinite-constraints} and \eqref{eq:directional-derivative-general}, the directional derivative $dG_j$ can be expressed as
\begin{equation}\label{eq:directional-derivative-constraint}
    dG_j(u;h) 
    = \int_{0}^{T}d\tau\ h(\tau)\frac{\delta G_j}{\delta u}(u,\lambda_j,\tau),\quad j=1,\ldots,4,
\end{equation}
where the explicit expressions for $\frac{\delta G_j}{\delta u}$ can be found in \eqref{eq:kkt-infinite-lagrangian}.
Then, the first equation in \eqref{eq:kkt-conditions} is given by  
\begin{equation}
    \label{eq:kkt-conditions-operator}
    \begin{split}
        &\int_{0}^{T}d\tau\ h(\tau)\left[\frac{\delta J}{\delta u}(u,\tau)+ \sum_{j=1}^{4}\frac{\delta G_j}{\delta u}(u,\lambda_j,\tau)\right]=0, \\& \hspace{2.5in} \text{ for all \zlcomment{square-integrable} } h \text{ on }[0,T].\\
    \end{split}
\end{equation}
The condition \eqref{eq:kkt-conditions-operator} can be reduced to an infinite-dimensional equation as follows. 
By the fundamental lemma of calculus of variations \citep[Lemma 3.2.3]{jost1998calculus}, \eqref{eq:kkt-conditions-operator} implies that the integrand is equal to zero for almost every $\tau\in[0,T]$.
Then, the KKT conditions \eqref{eq:kkt-conditions} reduce to the following infinite-dimensional system of equations and inequalities:
\begin{equation}\label{eq:kkt-conditions-infinite}
    \begin{split}
        &\frac{\delta J}{\delta u}(u,\tau) + \sum_{j=1}^{4}\frac{\delta G_j}{\delta u}(u,\lambda_j,\tau) = 0,\quad \zlcomment{\text{for almost every } \tau\in[0,T]}, \\
        &g_j(x,t,u)\leq 0\text{ and } \lambda_j(x,t)\geq 0,\quad (x,t)\in\Omega_j, \quad j=1,\ldots,4, \\
        &G_j(u,\lambda_j) = 0, \quad j=1,\ldots,4. \\
    \end{split}
\end{equation} 
Note that \eqref{eq:kkt-conditions-infinite} is written in the form \eqref{eq:kkt-infinite-dim}.

\bibliographystyle{plainnat}
\bibliography{references}

\end{document}